\documentclass[12pt]{spieman}  
\usepackage{amsmath,amsfonts,amssymb}
\usepackage{graphicx}
\usepackage{setspace}
\usepackage{tocloft}
\usepackage{float}

\makeatletter
\renewcommand*\env@matrix[1][*\c@MaxMatrixCols c]{%
  \hskip -\arraycolsep
  \let\@ifnextchar\new@ifnextchar
  \array{#1}}
\makeatother

\title{Hybrid propagation physics for the design and modeling of astronomical observatories: a coronagraphic example}

\author[a]{Jaren N. Ashcraft}
\author[b]{Ewan S. Douglas}
\author[a,b,c]{Daewook Kim}
\author[d]{A.J.E. Riggs}
\affil[a]{James C. Wyant College of Optical Sciences, University of Arizona, Meinel Building 1630 E. University Blvd., Tucson, AZ. 85721, USA}
\affil[b]{Department of Astronomy and Steward Observatory, University of Arizona, 933 N. Cherry Ave., Tucson, AZ 85721, USA}
\affil[c]{Large Binocular Telescope Observatory, University Of Arizona, 933 N. Cherry Ave. Tucson, AZ 85721, USA} 
\affil[d]{Jet Propulsion Laboratory, California Institute of Technology, 4800 Oak Grove Drive, Pasadena, CA 91109, USA}

\cftpagenumbersoff{figure}
\cftpagenumbersoff{table} 
\begin{document} 
\maketitle

\begin{abstract}
For diffraction-limited optical systems, an accurate physical optics model is necessary to properly evaluate instrument performance. Astronomical observatories outfitted with coronagraphs for direct exoplanet imaging require physical optics models to simulate the effects of misalignment and diffraction. Accurate knowledge of the observatory's point-spread function (PSF) is integral for the design of high-contrast imaging instruments and simulation of astrophysical observations. The state of the art is to model the misalignment, ray aberration, and diffraction across multiple software packages, which complicates the design process. Gaussian Beamlet Decomposition (GBD) is a ray-based method of diffraction calculation that has been widely implemented in commercial optical design software. By performing the coherent calculation with data from the ray model of the observatory, the ray aberration errors can be fed directly into the physical optics model of the coronagraph, enabling a more integrated  model of the observatory. We develop a formal algorithm for the transfer-matrix method of GBD, and evaluate it against analytical results and a traditional physical optics model to assess the suitability of GBD for high-contrast imaging simulations.Our GBD simulations of the observatory PSF, when compared to the analytical Airy function, have a sum-normalized RMS difference of $\approx 10^{-6}$. These fields are then propagated through a Fraunhofer model of an exoplanet imaging coronagraph where the mean residual numerical contrast is 4 $\times 10^{-11}$, with a maximum near the inner working angle at 5 $\times 10^{-9}$. These results show considerable promise for the future development of GBD as a viable propagation technique in high-contrast imaging. We developed this algorithm in an open-source software package and outlined a path for its continued development to increase the accuracy and flexibility of diffraction simulations using GBD.

\end{abstract}

\keywords{physical optics modeling, coronagraphs, exoplanets, high-contrast imaging, diffraction, Gaussian Beamlet Decomposition}

{\noindent \footnotesize\textbf{*}Jaren N. Ashcraft,  \linkable{jashcraft@arizona.edu}, Ewan S. Douglas, douglase@arizona.edu }

\begin{spacing}{2}   

\section{Introduction}
\label{sec:intro}  
\subsection{Astrophysical Motivation}

Integrated models of optical observatories are highly beneficial to their design and use\cite{andersen_integrated_2011,Dube2022}. Accurate observatory models permit powerful insight into predicting the as-built performance of a given instrument. However, the accuracy of these models is fundamentally limited by the assumptions made. To facilitate high-yield scientific observations, astronomical observatories are nominally designed to operate in the diffraction limit where wavefront aberrations are small. Diffraction-limited optical observatories are necessarily modeled with diffraction integrals derived from the Huygens-Fresnel principle to support the wave-like behavior of light. The paraxial and scalar assumptions that angles of incidence are small and that polarization is negligible\cite{goodman17} are made to ease the computational burden on the model. The resultant Fresnel and Fraunhofer diffraction integrals are accurate providing these conditions are met. If the performance of the observatory is limited by a factor outside the assumptions made, then the model will be ignorant of it. An example of this is the linear and shift-invariant assumption imposed on diffraction models of astronomical observatories. Ray aberrations (e.g. coma, astigmatism) have a field dependence, and consequently change across an observatory's field of view. However, diffraction integrals assume shift-invariance. This means that the aberrations do not change across the field of view and a separate ray trace model must be used to capture this effect.

Integrating optical models from different regimes in physics has become a popular method by which to overcome this limitation. Linking ray trace models to diffraction models in particular can overcome the paraxial and scalar assumption imposed by the Fresnel and Fraunhofer diffraction integrals. In the prior example, to capture the influence of optical aberrations, a new ray trace must be performed and the optical path difference of the rays must be translated to a diffraction model for each point of interest in the field of view. Similarly, diffraction integrals are incapable of determining the effects of optical polarization. For example, the Daniel K.~Inoyue Solar Telescope (DKIST) supports a suite of polarimetric instrumentation that is sensitive to the influence of optical polarization. To support this regime of optical physics the scalar assumption is not sufficient, so the polarization state is propagated along geometric ray paths using polarization ray tracing\cite{Chippman15,anche_inprep} to determine the influence of polarization aberrations on the optical beam\cite{Harrington2017}. Modern space telescopes also require integrated models to accurately predict the instrument behavior. For example, the optical models of the James Webb Space Telescope incorporate the influence of dynamic thermal, structural and optical effects simultaneously to produce an accurate library of the observatory's jitter\cite{Mather2004}. High-contrast imaging instruments that use coronagraphs, designed to separate exoplanets from diffracted starlight, are in dire need of integrated physical optics modeling from their inception. These coronagraphs aim to discern targets that are orders of magnitude dimmer than their host star\cite{Guyon_2006,2015ApJStark,2018Pueyo}. Understanding all sources of error is of paramount importance to the functionality of the instrument.

High-contrast imaging instruments have been successfully deployed on the ground (e.g. SCExAO\cite{Lozi18}, MagAO-X\cite{Males18}, NIRC2\cite{Femenia16}, GPI\cite{Macintosh14}, SPHERE\cite{refId0}) and in space (e.g. NICMOS\cite{thompson_near_1994}, NIRCam\cite{horner_near-infrared_2004,Nircam_exoplanet}) to pursue the direct detection of extrasolar planets, debris, and protoplanetary disks. 
The Decadal Survey on Astronomy and Astrophysics 2020 (Astro2020) recommends pursuing these instruments for a future 6 meter diameter infrared/optical/visible (IROUV) flagship observatory for the progression of astrophysical sciences\cite{decadal_survey_on_astronomy_and_astrophysics_2020_astro2020_pathways_2021}. 

Presently the optical design of observatories is done in a ray-tracing engine\cite{Howard22,Howard11} (e.g. CODE V, Zemax OpticStudio) because it is more suitable to optimizing the shapes of observatory mirrors. Upon reaching a diffraction-limited optical design, the system is then assumed to be well-represented by a paraxial diffraction model. The wavefront maps produced by the ray trace model of the observatory and the contributions from the imperfect polishing of the observatory mirrors are sent to a linearized physical optics propagator to examine the image plane electric field in the presence of diffraction from structure in the beam and phase errors on the optics. 

Many tools have been developed to simulate the performance of high-contrast imaging instrumentation. Tiny Tim is one of the first of these widely-used packages used to simulate the Hubble Space Telescope (HST)  instrument point-spread functions (PSFs)\cite{Krist93}. The tool generates aberrated PSFs based on the instrument, observation, and dynamic aberrations for a given observing scenario, enabling highly accurate simulations of the observatory performance. However, it only considers aberrations that are conjugate to the exit pupil of the observatory. This limits the model's ability to capture out-of-pupil effects, like the Talbot effect and speckles from optical surfaces\cite{goodman17}. To capture these effects, optical propagation packages integrate optical models of observatories by adding Fresnel diffraction to the PSF simulation, enabling the modeling of plane-to-plane diffraction effects. Open-source packages that currently support Fresnel diffraction include: PROPER \cite{Krist07}, Physical Optics Propagation in PYthon (POPPY)\cite{Perrin12,2016ascl.soft02018P,Doug18}, High-Contrast Imaging in Python (HCIPy) \cite{por2018hcipy}, AOTools\cite{Townson:19}, and prysm \cite{Dube2022,Dube2019}. Using these tools, near-field diffraction that limits high-contrast imaging can be modeled, and focal-plane wavefront sensing methods can be tested. 

These open-source physical optics propagation tools form the cornerstone of high-contrast imaging instrument modeling and design. The open-source framework means that the codes are accessible to anyone, so the physics are completely verifiable by the scientific community\cite{Allen2021OpenSource}. It is in the scientific community's best interest to continue to develop open-source propagation physics modules to increase the scope of and further integrate our observatory models. 

Commercial optical design codes offer the ability to make diffraction calculations based on ray data, but their physical optics simulation techniques are not as transparent or versatile as the open-source propagation codes that are used to design coronagraphs for astronomical observatories. The current open-source physical optics codes used for observatory modeling are also limited in their scope because of the Fresnel approximation, which is incapable of accurately modeling the field after fast-focusing and highly aspheric surfaces\cite{krist_practical_2010,vanderbei_diffraction_2006}. As observatories get larger, their optics may become faster (i.e. lower $F\#$) and more aspheric to fit within an available volume. Some coronagraph architectures capable of Earth-like exoplanet detection (e.g. phase-induced amplitude apodization, or PIAA \cite{guyon_phase_nodate}) employ mirrors that apodize the pupil with highly aspheric mirrors, and require tailored propagators in order to be included in physical optics models\cite{krist_practical_2010}. In the regime where the contribution of these surfaces is best represented by a ray trace, a diffraction calculation must be made to appropriately model the optical field at the image plane. To continue the development of integrated optical models, exploring the possibilities and limitations of new propagation techniques is desirable.

\begin{figure}[H]
	\centering
	\includegraphics[width=\textwidth]{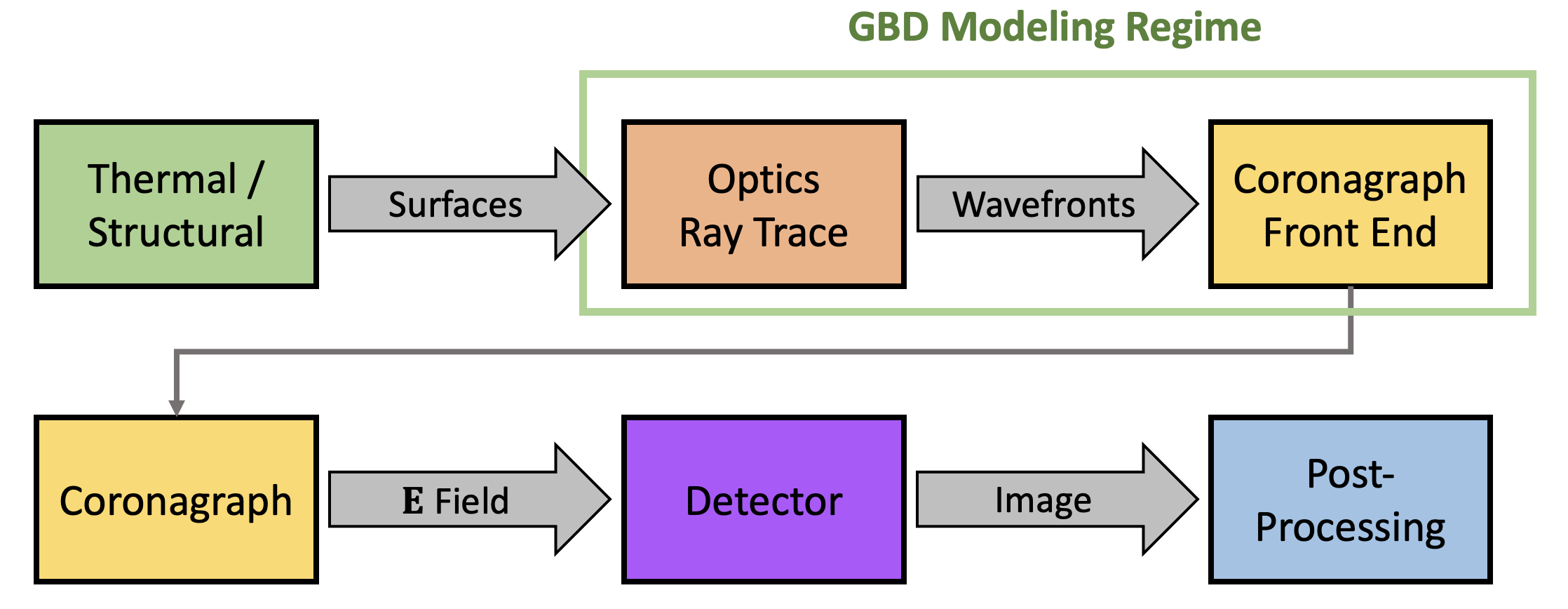}
	\caption{Modeling flow inspired by the Structural Thermal Optical Performance (STOP) modeling process for the Roman Coronagraph \cite{Krist18}. This diagram illustrates the different modeling regimes required to create a simulated image of an observatory. We aim to further integrate this modeling pipeline by creating an open-source Gaussian Beamlet Decomposition platform to unify the ray trace model of the observatory with the diffraction model of the coronagraph.}
    \label{fig:modeling_flow}
\end{figure}

An example of the typical integrated modeling pipeline for astronomical observatories outfitted with coronagraphs is shown in Figure \ref{fig:modeling_flow}. The observatory is typically designed and modeled in ray trace software (e.g. CODE V, Zemax OpticStudio) to accurately model the wavefront in the observatory's exit pupil. Upon finalizing the design, the complex-valued exit pupil is decomposed into a functional representation (e.g. a set of polynomial coefficients, such as the Zernikes\cite{krist_numerical_2015}) and passed to the entrance pupil of a coronagraph model constructed in an open-source, Fourier-based physical optics propagator (e.g. POPPY, PROPER, HCIPy). The front-end model computes the complex field distribution at the coronagraph mask, and then propagates the field past the mask to the image plane. The field is taken by a model of the detector (e.g. EMCCD Detect\cite{Nemati2023}, Pyxel\cite{Pyxel}) to create a simulated raw science image that can be post-processed (e.g. PyKLIP\cite{pyklip}, NMF imaging\cite{nmfimaging}).

To bridge the gap between commercial ray tracing engines and open-source physical optics propagation codes, we investigate the viability of a ray-based diffraction calculation called Gaussian Beamlet Decomposition (GBD) for modeling observatories with coronagraphs. Traditionally, GBD operates using the complex ray tracing algorithm described in the works by Greynolds\cite{Greynolds86} and Harvey et al\cite{Harvey15}. This technique has been previously implemented in FRED\cite{modeling_coherence_fred,Harvey15}, and possibly in CODE V\cite{bsp_in_codev}, but an exact method of its implementation in these software packages is not clearly available in the literature. An alternative approach called the transfer matrix algorithm was recently developed to improve GBD's viability for precision diffraction simulation by Worku and Gross\cite{Worku17,Worku:18,Worku19}. However, their implementation is not public and has not yet been formally evaluated as a tool to augment the modeling of astronomical observatories or high-contrast imaging instrumentation. To formally evaluate GBD as a modeling tool for astronomical instrumentation, a complete algorithm for its implementation is derived in this manuscript.

\subsection{Gaussian Beamlet Decomposition}
\begin{figure}[H]
    \centering
    \includegraphics[width=\textwidth]{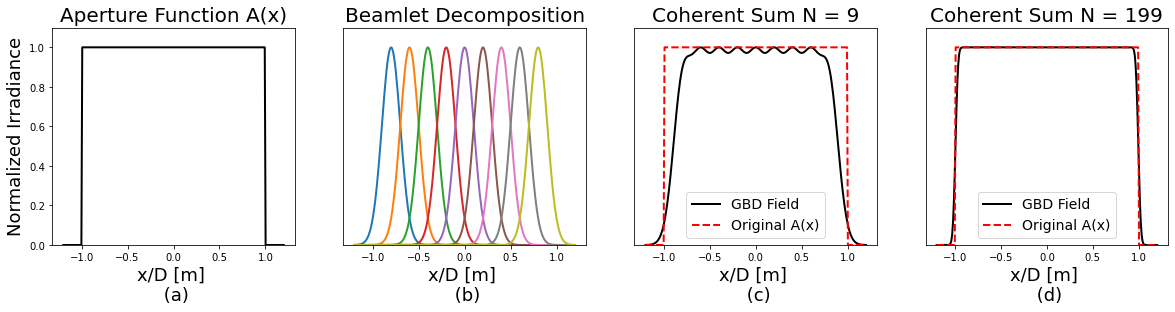}
    \caption{Illustration of the operating principle of GBD in one dimension. The aperture function for traditional imaging systems is shown in (a) as a top-hat function. The decomposition of this function into a discrete set of Gaussian beams is shown in (b), which shows nine evenly-spaced Gaussian profiles before their coherent summation, which is shown in (c). The coherent sum (black) of these nine beamlets shows that the beamlets are incapable of perfectly reconstructing the original aperture function (dashed red), specifically the sharp edge and uniform amplitude. More beamlets are needed for a more accurate reconstruction, which is shown in (d) for 199 beamlets. The amplitude ripple has virtually vanished, and the field near the aperture edges is almost entirely recovered.}
    \label{fig:gbd_essentials}
\end{figure}

GBD is a method of physical optics propagation that approximates the propagated field as a finite sum of coherent Gaussian beams that each propagate along a ray path. This method has been implemented in optical design packages \cite{greynolds_ten_2020} to perform coherent calculations on non-paraxial systems. The operating principle of GBD is to decompose the field in the entrance pupil of an optical system (Figure \ref{fig:gbd_essentials}a) into a finite set of Gaussian beams (Figure \ref{fig:gbd_essentials}b). Their coherent sum (Figure \ref{fig:gbd_essentials}c-d) approximates the initial field decomposition and can be propagated anywhere in the optical system along geometric ray paths.

Under-sampling the field in the entrance pupil leads to artifacts in the decomposition. A characteristic amplitude ripple based on the period of the beamlet decomposition remains in the field. Due to the soft edges of the Gaussian beams, they cannot completely reconstruct the field of a sharp aperture edge. Using a larger number of smaller beamlets decreases the beamlet distribution period and increases the slope of the Gaussian beams, allowing for the mitigation of both of these effects (shown in Figure \ref{fig:gbd_essentials}d).

Upon decomposing the initial wavefront into a sufficient set of Gaussian beams, we can use their analytical linear propagation laws\cite{Siegman_1986} (described in \hyperref[sec:methods]{Section \ref{sec:methods}} of this manuscript) to compute the coherent field at any arbitrary plane in the optical system. Fourier transform-based propagation methods derived from the Huygens-Fresnel principle typically assume the field is scalar and the optical system is paraxial \cite{goodman17}. This is usually appropriate for stellar coronagraphs, which operate on slowly focusing beams with diffraction-limited optics, but may not be for next-generation observatories whose large apertures may necessitate relatively fast telescope optics.

GBD  computes the same complex optical field without making the paraxial assumption across the observatory. Rather, the coherent field of  Gaussian beams is derived from the ray data directly. Doing so imposes the paraxial assumption about a single beamlet instead of the entire observatory, which is a much less stringent approximation. Gaussian beams are technically infinite in extent, but extremely localized around the beam waist. Consequently, the contribution of the field very far from the Gaussian is negligible. This locality enables the simulation of the optical system to generally be non-paraxial. By making the diffraction calculation directly from ray data, GBD circumvents the need for translating the wavefront to a physical optics propagator and imposing the paraxial assumption on the optical system. Instead, the ray trace model can be directly integrated into the diffraction model. 

There are two main approaches that exist in the literature to implement GBD: the complex ray tracing method, and the transfer matrix method. The complex ray tracing method was recently described by Harvey et al\cite{Harvey15} in their seminal paper about implementing GBD in Photon Engineering's non-sequential ray tracing software FRED. This method traces waist and divergence rays to compute the complex field at the plane of interest using Arnaud's method of complex ray tracing\cite{arnaud_representation_1985,Greynolds86}. Through FRED, the complex ray tracing method has seen widespread use for nonparaxial coherent beam analysis.

Another GBD approach, the transfer matrix method was developed by Worku and Gross\cite{worku_vectorial_2017,Worku:18,Worku19} to mitigate GBD's inability to simulate sharp-edge diffraction and add new utility to the technique. This formulation of GBD works by computing the differential ray transfer matrix for a given ray path and then using that data to solve the Gaussian beam solution to the general Collins integral\cite{Collins:70}. Worku and Gross have leveraged the general Collins integral to provide alternative conditions to the Gaussian beam solution to modify the decomposition, such as truncated\cite{Worku19} and pulsed\cite{Worku:20} beamlet decomposition. The option of modifying the  beamlets to overcome the limitations of GBD makes the transfer matrix method extremely attractive for use in high-contrast imaging where preservation of high-spatial frequency content is important. Of particular interest are mirror segment gaps and opto-mechanical structures that obscure the primary mirror. Therefore, we elect to investigate Worku and Gross's transfer matrix method of GBD for the work presented in this manuscript. Our goal is to publicize the transfer matrix method by developing an algorithm for its implementation, and then use it to characterize GBD's suitability for high-contrast imaging simulation. Our work can then be used as a platform with which to study the suitability of Worku and Gross' modified GBD in future investigations.

\subsection{Hybrid Propagation Physics}
\begin{figure}[H]
    \centering
    \includegraphics[width=\textwidth]{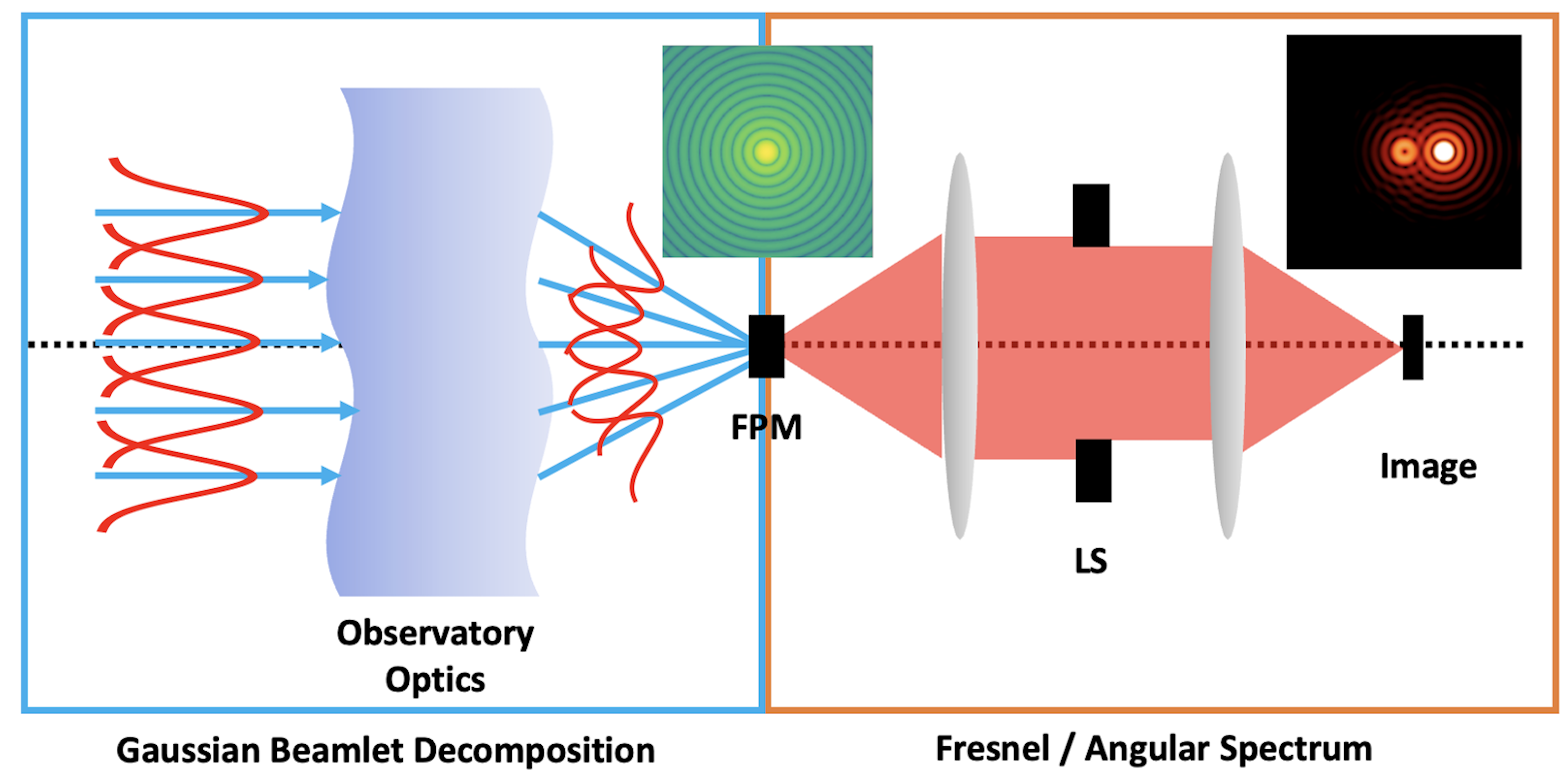}
    \caption{Diagram demonstrating the hybrid propagation physics model. The observatory optics are best described by a ray-based propagation model, so GBD is used to compute the field at the image plane before the coronagraph focal plane mask (FPM). The field array is then passed to a paraxial diffraction model which propagates it past the FPM and through the remainder of the coronagraph. This propagation scheme permits the user to model the influence of the observatory optics non-paraxially, without losing accuracy after propagating past the focal plane mask.}
    \label{fig:hybridpropdiagram}
\end{figure}

The ray-based nature of GBD introduces problems in modeling the electric field when the rays are vignetted. Structure in the field where the initial decomposition occurs (typically, the entrance pupil) can be well-represented by Gaussian beams as long as the structure of interest is larger than the beamlets used in decomposition. Diffraction from structure in interemediate planes (between the pupil and focal planes) is challenging to represent if the beamlets diverge considerably. However, secondary beamlets can be traced from these intermediate structures to aid in the accuracy of the simulation\cite{Greynolds14}. At the focal plane of a diffraction-limited system, the rays are highly concentrated while the diffracted field spreads out considerably (e.g. the Airy disk). Because of GBD's reliance on ray tracing, it cannot represent diffraction from structure in the focal plane well without re-decomposing the field \cite{modeling_coherence_fred}. For the case of a Lyot-type coronagraph all rays are vignetted at the focal plane mask (FPM) and the field decomposition is lost. To circumvent this we compute the field before the FPM with GBD and propagate it through the remaining coronagraph with traditional diffraction integrals, where we expect the low-order aberrations to be small and the paraxial assumption to be valid. This \emph{hybrid} method (shown in Figure \ref{fig:hybridpropdiagram}) enables the user to alter the propagation physics for the electric field based on where it is the most appropriate; GBD will simulate the fast beams in the fore-optics and paraxial diffraction will propagate the field through the coronagraph imaging optics. In practice, GBD would be used to propagate to the entrance pupil of the coronagraph for simulating systems with wavefront control, where it would then hand off the field to a paraxial diffraction model.

The end result of such a model allows for direct integration of the ray trace model with the physical optics model, without imposing the paraxial approximation on the observatory. Like Fresnel and Fraunhofer diffraction, GBD is an approximation to diffraction physics. The decomposition of the field into Gaussian beams does not have an analytical solution. Therefore, undesirable artifacts can be introduced into the field if the decomposition is not well-understood and the sampling is insufficient\cite{Ashcraft2020}. To better understand the impact of a GBD PSF on high-contrast imaging simulations, we develop a hybrid propagation model to compare GBD to an equivalent Fraunhofer diffraction model.

To our knowledge, the transfer matrix method of GBD has not seen widespread implementation. Given its obvious benefits, we believe that this is because the transfer matrix method is not well-understood by the scientific community. We aim to remedy this by presenting a vectorizeable algorithm for the transfer matrix method, and open-sourcing our simulation platform for future investigators to use. This manuscript is the first work, to our knowledge, to provide the explicit mathematics of GBD \emph{and} provide our code as an object-oriented module for more widespread use. 

Our GBD module was built in the Poke\cite{Ashcraft_poke_2022} Python package currently available on Github. Poke was originally developed to be a polarization ray tracing module to study the influence of polarization aberrations on astronomical coronagraphs\cite{anche_inprep}. For this study, we expanded its capabilities to include GBD. Poke operates by using ray tracer API's to trace a raybundle through every surface in the optical system. The relevant ray data is stored in a Rayfront object and can be loaded into a Python environment and interacted with independent of the ray tracer that generated the data. The Rayfronts can also be compiled into binary file types using the msgpack\cite{msgpack} package and distributed to any interested investigator, effectively open-sourcing the physical optics calculations done on ray data.
\begin{figure}[H]
    \centering
    \includegraphics[width=0.9\textwidth]{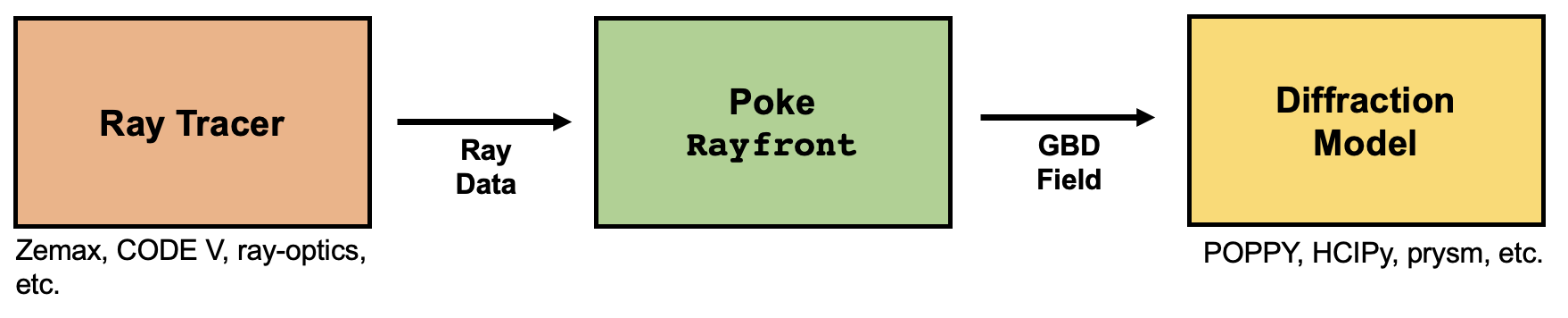}
    \caption{Illustration of Poke's use as an interface to open-source ray-based physical optics. Poke only requires an interface between a ray tracing engine (orange, left) to generate and save a Rayfront object, which includes all ray data necessary for the GBD calculation. Currently, Poke supports sequential systems in CODE V and Zemax, but we are working on adding support for open-source packages that have ray tracers, like ray-optics\cite{rayoptics}. The GBD field that Poke computes can then be sent to any open-source diffraction modeling package (yellow, right) to complete the hybrid propagation model.}
    \label{fig:poke_modeling}
\end{figure}

In this study we conduct ray traces in Zemax, which are then saved as a Poke Rayfront object. Poke performs the GBD simulations using the saved ray data to generate the field at the focal plane of the telescope. This data is exported to a coronagraph model built using HCIPy.

In \hyperref[sec:methods]{Section \ref{sec:methods}} we outline the mathematics of Gaussian beam propagation and differential ray tracing used to perform GBD simulations. In \hyperref[sec:algorithm]{Section \ref{sec:algorithm}} we present the mathematical algorithm for the transfer-matrix method of GBD that we developed for this investigation. In \hyperref[sec:results]{Section \ref{sec:results}} we compare the results of observatory PSFs produced by GBD with one produced using traditional diffraction methods, and analyze the artifacts that remain in the field. In \hyperref[sec:conclusion]{Section \ref{sec:conclusion}} we assess the suitability of GBD for high-contrast imaging models and establish a roadmap for our module's continued development. 

\section{Preliminary Mathematical Methods}
\label{sec:methods}  

In this Section we review the necessary mathematical tools that we need to formulate our GBD algorithm. This includes the propagation equations for a single Gaussian beam, how to use differential ray tracing to compute the parameters necessary for Gaussian beam propagation, and methods of decomposing the entrance pupil field in the optical system to improve the simulation's sensitivity to high-spatial frequencies. 

\subsection{Propagation of a single Gaussian beam}
The equation for a single Gaussian Beam is parameterized entirely by the complex beam parameter $Q(z)$, \cite{goodman17}
\begin{equation}
	U(r,z) = \frac{U_o}{Q(z)}exp\Bigl[ik\frac{r^2}{2Q(z)}\Bigr],
\end{equation}
where $U(r,z)$ is the scalar Gaussian field, $U_o$ is the amplitude, $k$ is the wavenumber, and $r$ is the radial coordinate in the plane perpendicular to propagation. The inverse of the complex parameter $Q(z)$ describes the beam's $1/e$ field radius (the ``waist" $w(z)$) and wavefront radius of curvature $R(z)$,
\begin{equation}
	Q(z)^{-1} = \frac{1}{R(z)}+i\frac{\lambda}{\pi w(z)^2}.
\end{equation}	
$Q(z)^{-1}$ is a convenient expression of the Gaussian beam because it fully encapsulates the information required to describe the transverse electric field of the beam as it propagates. The real part of $Q(z)^{-1}$ is related to the radius of curvature of the wavefront,
\begin{equation}
    R(z) = z\Bigl(1+(\frac{Z_o}{z})^2\Bigr),
    \label{eq:complex_curvature}
\end{equation}
where $Z_o$ is the Rayleigh range and $z$ is the longitudinal propagation distance.
The imaginary part of $Q(z)^{-1}$ is related to the beam waist radius,
\begin{equation}
	w(z) = w_o\sqrt{1+\Bigl(\frac{z}{Z_o}\Bigr)^2}.
\end{equation}
In the paraxial regime $Q(z)^{-1}$ can be propagated using the ABCD ray transfer matrices of geometrical optics\cite{Siegman_1986}.
\begin{equation}
    Q(z)^{-1}_{2} = \frac{C + DQ_{1}^{-1}}{A + BQ_{1}^{-1}}
    \label{eq:propq}
\end{equation}
To account for system misalignments, $Q(z)^{-1}$ is a 2x2 matrix $\textbf{Q}(z)^{-1}$ that encodes the complex curvature in two orthogonal directions and how they couple into each other, allowing for the beamlet to be generally astigmatic.\cite{Ashcraft2020,cai_decentered_nodate}. 

\begin{equation}
    \textbf{Q}(z)^{-1} = 
    \begin{pmatrix}
    Q(z)_{xx}^{-1} & Q(z)_{xy}^{-1} \\
    Q(z)_{yx}^{-1} & Q(z)_{yy}^{-1} \\
    \end{pmatrix}
\end{equation}

This treatment allows for greater versatility in the beamlet propagation, but requires that the elements of the ray transfer matrices are also 2x2 matrices. The propagation formula shown in Equation \ref{eq:propq} has a similar matrix extension.

\begin{equation}
    \mathbf{Q}(z)_{2}^{-1} = (\mathbf{C} + \mathbf{DQ}^{-1}_{1})(\mathbf{A} + \mathbf{BQ}^{-1}_{1})^{-1}
    \label{eq:prop_complex}
\end{equation}

We can then express the propagated Gaussian beam with Equation \ref{eq:gaussian_prop}
\begin{equation}
    U(\mathbf{r}) = \frac{U_{o}}{\sqrt{det|\mathbf{A} + \mathbf{BQ_{1}^{-1}}|}} exp\Bigl[\frac{-ik}{2} \mathbf{r}^{T} \mathbf{Q_{2}^{-1}} \mathbf{r}\Bigr],
    \label{eq:gaussian_prop}
\end{equation}

Where $\mathbf{r}$ is the radial coordinate in the plane transverse to the propagation direction centered on the Gaussian beam. The formulation for the propagation of the complex curvature matrix allows for modeling Gaussian beams as they propagate along generally skew ray paths in non-axially symmetric optical systems. This is an important utility for diffraction modeling of wavefront aberrations introduced by system misalignment or thermal deformations, which generally break optical system symmetry. Note that the solution in Equation \ref{eq:gaussian_prop} is only valid for propagation between planes that are orthogonal to the propagation direction of the Gaussian beam \cite{Collins:70}. We next need to determine a method for computing the ray transfer matrix for an arbitrary ray path through an optical system in order to propogate the Gaussian beam.

\subsection{Computing the Differential Ray Transfer Matrix}
The ABCD ray transfer matrix is a useful and concise method for analyzing properties of ray paths along optical systems. In the regime of geometrical optics, a generally skew ray can be traced through a system using 4x4 ABCD ray transfer matrices\cite{Brouwer64}. These matrices model simple optical elements (e.g. thin lenses) with ease by operating on an input column vector that represents a light ray. The simplest ray transfer matrix that describes a paraxial and orthogonal optical system is a 2x2 operator that maps an input ($i$) spatial and angular coordinate to the appropriate output ($o$),
\begin{equation}
    \begin{pmatrix}
    y_o \\
    m_o \\
    \end{pmatrix}
    =
    \begin{pmatrix}
    A & B \\
    C & D \\
    \end{pmatrix}
    \begin{pmatrix}
    y_i \\
    m_i \\
    \end{pmatrix},
\end{equation}
where $y$ is the spatial coordinate transverse to the propagation direction, and $m$ is the slope in that dimension. The elements of the ABCD matrix and ray vectors are real-valued scalars. To account for skew ray paths, the position and angle in the dimension orthogonal to $y$ and the direction of propagation must be tracked, adding two dimensions to the matrix calculus. A 4x4 ABCD matrix describes a nonorthogonal system with tilts and decenters that map generally skew input rays to generally skew output rays, 

\begin{equation}
    \begin{pmatrix}
    x_o \\
    y_o \\
    l_o \\
    m_o \\
    \end{pmatrix}
    = 
    \begin{pmatrix}
    A_{xx} & A_{xy} & B_{xx} & B_{xy} \\
    A_{yx} & A_{yy} & B_{yx} & B_{yy} \\
    C_{xx} & C_{xy} & D_{xx} & D_{xy} \\
    C_{yx} & C_{yy} & D_{yx} & D_{yy} \\
    \end{pmatrix}
    \begin{pmatrix}
    x_i \\
    y_i \\
    l_i \\
    m_i \\
    \end{pmatrix}.
    \label{eq:totalabcdmatrix}
\end{equation}

For simplicity, it is convenient to represent the radial position in the plane transverse to propagation ($x$, $y$) and the corresponding direction in the dimension ($l$, $m$) as a position and angle vector respectively ($\mathbf{r}, \boldsymbol{\theta}$). The ABCD matrix can similarly be condensed into 2x2 sub matrices that operate on each spatial dimension, yielding a familiar notation,
\begin{equation}
    \begin{pmatrix}
    \mathbf{r_{o}} \\
    \boldsymbol{\theta_{o}} \\
    \end{pmatrix}
    =
    \begin{pmatrix}
    \mathbf{A} & \mathbf{B} \\
    \mathbf{C} & \mathbf{D} \\
    \end{pmatrix}
    \begin{pmatrix}
    \mathbf{r_{i}} \\
    \boldsymbol{\theta_{i}} \\
    \end{pmatrix}
    \label{eq:matrixabcd}.
\end{equation}

This description is powerful because it communicates the elegance and simplicity of ray transfer matrices. All dimensions transverse to propagation are accounted for, but the calculus to propagate a ray is still the same. The ray transfer matrices for simple and paraxial optical elements (e.g. thin lens) are well known\cite{Brouwer64}, and were used in concert with GBD in a prior investigation with paraxial systems\cite{Ashcraft2020}. However, our aim is to use GBD to model \emph{non}-paraxial optical system diffraction. Therefore, we need a method of computing the ray transfer matrix for an arbitrary skew ray-path. 
A simple dimensional analysis of the matrix relation in Equation \ref{eq:matrixabcd} is a good place to start understanding how to construct an arbitrary ABCD matrix. Because the position element $\mathbf{r}$ of the ray vector must be in units of distance, and the angular element $\boldsymbol{\theta}$ must be dimensionless, the units of the ABCD matrix elements are constrained. $\mathbf{A}$ and $\mathbf{D}$ must be dimensionless and transform the ray position and angle through the optical system, indicating that they represent magnification. $\mathbf{B}$ and $\mathbf{C}$ must have units of distance and inverse distance, respectively. $\mathbf{B}$ operates on an angle, and is therefore a metric of propagation distance through an optical system given some ray angle. $\mathbf{C}$ operates on a position, and is therefore an indicator of the amount of refraction a ray experiences given a position in the entrance pupil.

Stone and Forbes'\cite{Stone:97} work in differential ray tracing for inhomogeneous media was instructive in terms of deriving a method to construct the ABCD matrix. They illustrate the construction of individual optical elements through ray derivatives in a generally 4x4 matrix through derivatives of surface data. Using this method the position and angular derivatives of an optical surface are taken and arranged in a matrix like in Equation \ref{eq:matrixabcd}. The matrix product of the optical elements is then the final differential ray transfer (or ABCD) matrix. Their method is functional if the analytical expression of the optical elements are known, but could be very computationally intensive if the optical system has many elements. Instead, we approximate the differential matrix by tracing additional rays, and compute the finite difference of the ray coordinates and directions at the input and output of the optical system to approximate the derivative.

Our implementation of the differential ray tracing technique in Poke utilizes a user-specified ray tracer (e.g. Zemax OpticStudio) that propagates rays by computing Snell's law at each surface. For GBD, the ray coordinates of interest are on the source plane (where the field decomposition is done, e.g. the entrance pupil) and the transversal plane. The transversal plane is the plane normal to the central ray of a Gaussian beam that intersects the point at which we wish to evaluate the field. Propagation to this plane is critical, because the solution to Gaussian beam propagation (Equation \ref{eq:gaussian_prop}) is only valid between planes orthogonal to the propagation direction defined by the central ray. An element of the ray transfer matrix can be computed by determining the ratio of the differential ray data on the transversal plane to the differential ray data on the source plane. An example of computing the element $A_{yy}$ is given by Equation \ref{eq:Ayy}, 

\begin{equation}
    A_{yy} = \frac{\partial y_{T}}{\partial y_{S}} = \frac{y_{+y,T} - y_{cen,T}}{y_{+y,S} - y_{cen,S}}.
    \label{eq:Ayy}
\end{equation}

\begin{figure}[H]
    \centering
    \includegraphics[width=0.8\textwidth]{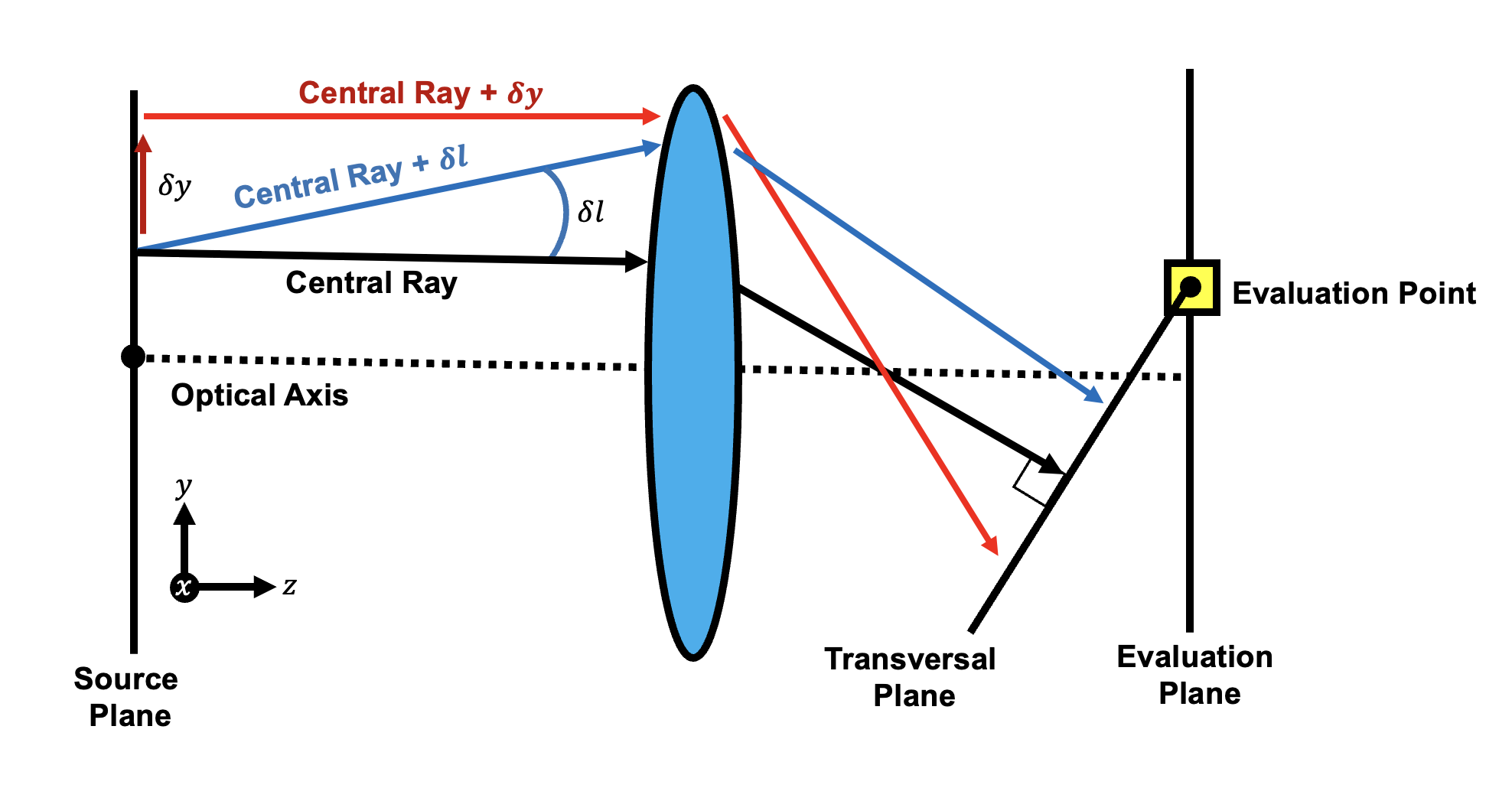}
    \caption{Diagram illustrating differential ray tracing in the 2D case for a simple lens system. The central ray (black) is propagated along with two parabasal rays with a differential addition in the $y$ direction (red) and another with a differential addition in the $l$ direction (blue). To determine the ABCD matrix the ray data is computed on the transversal plane, which is normal to the central ray and intersects the point at which we wish to evaluate the field.}
    \label{fig:diffdiagram}
\end{figure}

Where $y_{+y,T}$, and $y_{cen,T}$ are the ray coordinates of the differential ray and the central ray, respectively, on the transversal plane. $y_{+y,S}$, and $y_{cen,S}$ are the coordinates of the same rays on the source plane.  In this example the central ray (shown in black on Figure \ref{fig:diffdiagram}) is traced along with a ray with a differential addition in input $y$ coordinate (shown in red on Figure \ref{fig:diffdiagram}). The difference in $y$ coordinates of the central and $\delta y$ differential ray determine the derivative. 

The ray transfer matrix for a non-orthogonal optical system has 16 unknowns (see ABCD matrix in Equation \ref{eq:totalabcdmatrix}), and each ray yields 4 quantities. To solve for every element of the matrix 4 linearly independent rays must be traced. The simplest ray set is geometrically orthogonal\cite{Greynolds86}, where copies of the central ray $(x,y,l,m)$ are modified by a differential quantity $(\delta)$ in each of the 4 ray coordinates, two in position ($\delta x$, $\delta y$) and two in slope ($\delta l$, $\delta m$). The differential ray set is given by Equation \ref{eq:diffrays}
\begin{center}
    
\begin{equation}
    \begin{pmatrix}
    x + \delta x \\
    y \\
    l \\
    m \\
    \end{pmatrix}
    ,
    \begin{pmatrix}
    x \\
    y + \delta y \\
    l \\
    m \\
    \end{pmatrix}
    ,
    \begin{pmatrix}
    x \\
    y \\
    l + \delta l\\
    m \\
    \end{pmatrix}
    ,
    \begin{pmatrix}
    x \\
    y \\
    l \\
    m + \delta m\\
    \end{pmatrix},
    \label{eq:diffrays}
\end{equation}

\end{center}
which are traced in addition to the central ray of interest. The full differential ray transfer matrix is given by Equation \ref{eq:diffmat_rays}. The ray transfer matrix is purely a function of the Cartesian position of the ray ($x, y$) and  slope of the ray in those directions ($l, m$) at the input and output of the optical system. An example of propagating a ray from the source plane to the transversal plane is shown in Equation \ref{eq:diffmat_rays}. Here the subscript $S$ refers to the coordinate on the source plane, and the subscript $T$ refers to the coordinate on the transversal plane. The elements of Equation \ref{eq:diffmat_rays} are computed in a similar fashion to the example in Equation \ref{eq:Ayy}, but with different ray data.

\begin{center}
\begin{equation}
    \begin{pmatrix}
    x_{T} & y_{T} & l_{T} & m_{T} \\
    \end{pmatrix}
    =
    \begin{pmatrix}
    \frac{\partial x_T}{\partial x_S} & \frac{\partial x_T}{\partial y_S} & \frac{\partial x_T}{\partial l_S} & \frac{\partial x_T}{\partial m_S}  \\
    \frac{\partial y_T}{\partial x_S} & \frac{\partial y_T}{\partial y_S} & \frac{\partial y_T}{\partial l_S} & \frac{\partial y_T}{\partial m_S}  \\
    \frac{\partial l_T}{\partial x_S} & \frac{\partial l_T}{\partial y_S} & \frac{\partial l_T}{\partial l_S} & \frac{\partial l_T}{\partial m_S} \\
    \frac{\partial m_T}{\partial x_S} & \frac{\partial m_T}{\partial y_S} & \frac{\partial m_T}{\partial l_S} & \frac{\partial m_T}{\partial m_S} \\
    \end{pmatrix}
    \begin{pmatrix}
    x_S \\
    y_S \\
    l_S \\
    m_S \\
    \end{pmatrix}
    \label{eq:diffmat_rays}
\end{equation}
\end{center}
 
With differential ray tracing we are able to propagate a single Gaussian beam through an arbitrary optical system using the ABCD matrix. To perform GBD, we next need to understand how to decompose the field in the entrance pupil of the optical system.

\subsection{Entrance Pupil Spatial Decomposition}
The final variable to constrain in GBD is how to appropriately decompose the field in the entrance pupil into a finite set of Gaussian beams. This problem is illustrated in 1D earlier in Figure \ref{fig:gbd_essentials}; however, for imaging systems the decomposition is a 2D problem. The fundamental Gaussian mode does not represent a complete set\cite{koshel_novel_2001}, and therefore the decomposition of the field is not unique. We must carefully consider how the beamlets are distributed in the entrance pupil for accurate diffraction calculations. 

\begin{figure}[H]
    \centering
    \includegraphics[width=\textwidth]{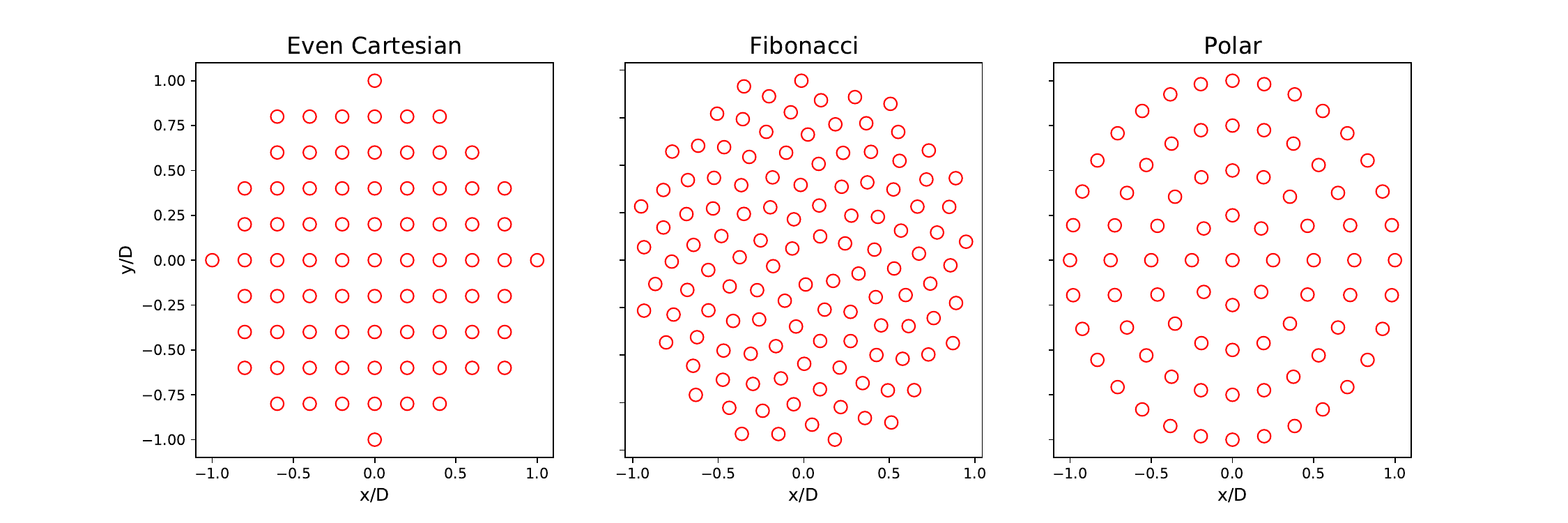}
    \caption{Illustration of the Even (left), Fibonacci (middle), and Polar (right) sample schemes for decomposing the field at the entrance pupil of the optical system with approximately the same number of beamlets in each Figure. Quantifying the ramifications of these sampling schemes is of paramount importance to accurate diffraction simulation.}
    \label{fig:sampleschemes}
\end{figure}

Various sampling schemes exist in the literature, with different strengths and weaknesses. The \emph{even Cartesian} sampling scheme (shown on the left in Figure \ref{fig:sampleschemes}) described in Harvey et al\cite{Harvey15} is the most straightforward, where the beamlets lie evenly spaced along a Cartesian grid. The ray coordinates in the entrance pupil are then computed from an overlap factor (OF) which describes the overlap of the beamlets' $1/e$ waist radii $w_o$,
\begin{equation}
    OF = \frac{N_{g} 2 \omega_{o}}{W},
\end{equation}
where $N_{g}$ is the number of Gaussian beamlets across an aperture, and $W$ is the width of the aperture. This feature is easy to implement and understand, but for under-sampled cases it introduces artifacts due to the ripple from the distribution and soft edge left by the Gaussian beamlets. 

The \emph{Fibonacci} sampling scheme (shown in the middle in Figure \ref{fig:sampleschemes}) introduced by Worku and Gross places the beamlets along a Fibonacci spiral, which results in a more accurate decomposition for circular apertures\cite{Worku:18}. The distribution of the beamlets is even along polar angles on the spiral. The polar distribution of the beamlets is given by a position $R$ and angle $\Theta$,
\begin{equation}
    R = \frac{W}{2}\sqrt{N_{g}},
\end{equation}
\begin{equation}
    \Theta = \frac{2\pi}{\phi^{2}}N_{g},
\end{equation}
where $\phi$ is the golden ratio and $N_{g}$ is the total number of beamlets to trace. Even polar sampling (shown on the right in Figure \ref{fig:sampleschemes}) is also a viable method to increase the accuracy of the decomposition for fewer beamlets assuming the optical system has a circular aperture\cite{Worku:18}, but was not explored in this study due to the apparent advantages of the Fibonacci sample scheme, which are shown in \hyperref[sec:results]{Section \ref{sec:results}}.

\section{The Proposed Beamlet Propagation algorithm}
\label{sec:algorithm}
\begin{figure}[H]
    \centering
    \includegraphics[width=\textwidth]{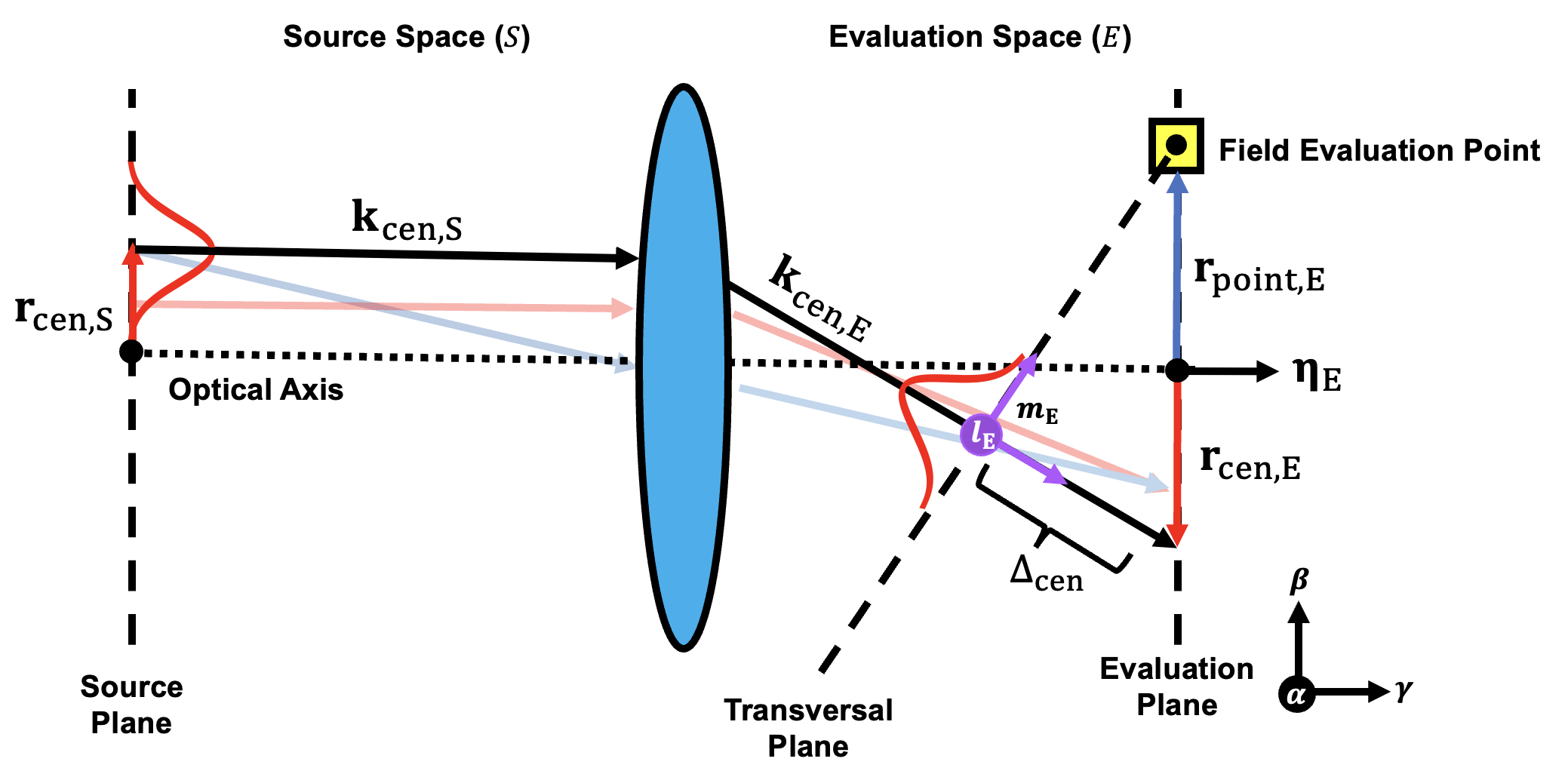}
    \caption{Diagram illustrating the geometry of GBD and the relevant data used in the propagation calculation for a single beamlet (shown in red). GBD begins by tracing the central ray (black) and differential rays (light red and blue) from the source plane (left) to the evaluation plane (right). The rays must be propagated from the evaluation plane to the transversal plane, which is normal to the central ray and intersects a point at which the field is evaluated (yellow box). The difference of the differential ray data with the central ray data on the transversal plane gives us the ABCD matrix used to propagate a Gaussian beamlet. We evaluate the Gaussian beamlet at the intersection of the transversal plane with the evaluation plane to get the beamlet's contribution to the total field at that point. This process is repeated for each point and each beamlet. The coherent superposition of the beamlets at the field evaluation points yields the final propagated field.}
    \label{fig:gaussian_prop_diagram}
\end{figure}

The transfer matrix method of GBD is the preferred algorithm to develop because of the recent developments by Worku and Gross to include modified GBD\cite{Worku19}. However, because the implementation is not public we must derive the full algorithm using the concepts described in \hyperref[sec:methods]{Section \ref{sec:methods}}. The basic concept of propagating a single beamlet through an arbitrary optical system is illustrated in Figure \ref{fig:gaussian_prop_diagram}. A Gaussian beam is placed at some position $\mathbf{r}_{cen,S}$ in the source plane where the initial decomposition occurs. The central ray (shown in black on Figure \ref{fig:gaussian_prop_diagram}) that tracks the position of the Gaussian beam and the differential rays (shown in light red and blue on Figure \ref{fig:gaussian_prop_diagram}) that define the propagation are traced to the evaluation plane using a ray tracing engine. The central ray position ($\mathbf{r}_{cen,E}$), and direction ($\mathbf{k}_{cen,E}$) on the evaluation plane are used to define the transversal plane which includes the point where we wish to evaluate the Gaussian field. The differential rays and the central ray are transformed to the transversal plane in order to compute the ABCD matrix that describes the propagation of the Gaussian beam from the source plane to the transversal plane. We use this matrix to compute the influence of this beam on the field evaluation point. Every beamlet used in the initial field decomposition is propagated to each point on the evaluation plane this way, and the coherent superposition of the beamlets at the evaluation plane represents the propagated field.

In this section we derive the formula to compute the propagation of a ray to the transversal plane, and how to use this data to compute the ABCD matrix and the contribution of a Gaussian beamlet to a point on the evaluation plane using the methods described in Section \ref{sec:methods}. We refer to several vectors in the algorithm below that are expressed using the convention $\mathbf{u}_{v,W}$. $\mathbf{u}$ is the data type of the vector, typically $\mathbf{r}$ if a position or $\mathbf{k}$ if a direction. $v$ denotes what item the vector belongs to, "cen" means it belongs to the central ray and "point" means a point on the evaluation plane. $W$ refers to the coordinate system of the vector, $S$ for "source space", $E$ for "evaluation space", or $T$ for "transversal plane". The relevant parameters used in the propagation algorithm are shown in Figure \ref{fig:gaussian_prop_diagram}, and are referenced throughout the procedure below.

\subsection{Propagating rays to the transversal plane}
A GBD simulation begins by running a ray trace though an optical system in the user's preferred design code. GBD needs the ray data at the plane where the decomposition occurred (the source plane) and the plane where we choose to observe the field (the evaluation plane). We use the ray data in evaluation space to propagate the rays to the transversal plane, where Equation \ref{eq:gaussian_prop} valid.

We first need to derive the propagation distance $\Delta_{ray}$ for the central and four differential rays. To do so, we find the intersection of the line defined by the ray we want to propagate $\mathbf{k}_{ray,E}$ and the plane normal to the central ray of the Gaussian beam $\mathbf{k}_{cen,E}$ which intersects the evaluation point $\mathbf{r}_{point,E}$. This plane is the transversal plane, and is defined by Equation \ref{eq:plane}

\begin{equation}
    \mathbf{k}_{cen,E} \cdot (\mathbf{r} - \mathbf{r}_{point,E}) = 0.
    \label{eq:plane}
\end{equation}

The line along the ray is defined by Equation \ref{eq:line}

\begin{equation}
    \mathbf{r} = \mathbf{r}_{ray,E} + \mathbf{k}_{ray,E}\Delta_{ray},
    \label{eq:line}
\end{equation}

where $\cdot$ denotes the dot product and $\mathbf{r}$ is the space of all points that satisfy Equations \ref{eq:plane} and \ref{eq:line}. To find the distance a ray needs to propagate along its own path in free space to intersect the transversal plane, we substitute $\mathbf{r}$ in Equation \ref{eq:plane} for Equation \ref{eq:line} and solve for $\Delta_{ray}$. The result is given by Equation \ref{eq:delta}

\begin{equation}
    \Delta_{ray} = - \frac{\mathbf{k}_{cen,E} \cdot (\mathbf{r}_{ray,E} - \mathbf{r}_{point,E})}{\mathbf{k}_{cen,E} \cdot \mathbf{k}_{ray,E}}.
    \label{eq:delta}
\end{equation}

The expression in Equation \ref{eq:delta} is written for the general case of propagating a ray from the evaluation plane to the transversal plane. It is applied to the central and differential rays by substituting $\mathbf{r}_{ray,E}$ and $\mathbf{k}_{ray,E}$ for the position and direction associated with the ray to propagate. We update the ray positons by invoking Equation \ref{eq:line}. The ray position on the transversal plane ($\mathbf{r}'_{ray,E}$) is given in Equation \ref{eq:prop_ray}

\begin{equation}
    \mathbf{r}'_{ray,E} = \mathbf{r}_{ray,E} + \mathbf{k}_{ray,E}\Delta_{ray}.
    \label{eq:prop_ray}
\end{equation}

However, these coordinates are still expressed in the basis of the evaluation plane. We must next rotate the ray coordinates into the transversal plane coordinate system so that they are orthogonal to the propagation direction.

\subsection{Transformation to the transversal plane}
In the evaluation plane coordinate system, we define the orthogonal basis vectors $\boldsymbol{{\alpha}}$,$\boldsymbol{{\beta}}$,and $\boldsymbol{{\gamma}}$ (shown in black on Figure \ref{fig:gaussian_prop_diagram} to the right of the evaluation plane), to be the directions along the $x$,$y$, and $z$ axes respectively. Similarly, the transversal plane is defined by basis vectors $\mathbf{l}_{E}$,$\mathbf{m}_{E}$,$\mathbf{k}_{cen,E}$, (shown in purple on Figure \ref{fig:gaussian_prop_diagram}) which are the analogous $x$,$y$, and $z$ directions for the transversal plane. To compute these directions, we start by taking the cross product of the central ray $\mathbf{k}_{cen,E}$ with the surface normal of the evaluation plane $\boldsymbol{\eta_{E}}$ (shown in black on the right of Figure \ref{fig:gaussian_prop_diagram})

\begin{equation}
    \mathbf{l}_{E} = \mathbf{k}_{cen,E} \times \mathbf{\eta_{E}}.
    \label{eq:compute_lbasis}
\end{equation}

A feature of Equation \ref{eq:compute_lbasis} is that the $\mathbf{l}_{E}$ is orthogonal to both the central ray and the surface normal. Similarly, to determine our final basis vector which is mutually orthogonal to both the central ray and $\mathbf{l}_{E}$ basis vector we compute their cross product

\begin{equation}
    \mathbf{m}_{E} = \mathbf{k}_{cen,E} \times \mathbf{l}_{E}.
    \label{eq:compute_mbasis}
\end{equation}

These vectors form a complete basis that describe the coordinate system of the transversal plane, but are expressed in the coordinate system of the evaluation plane. Therefore, we can construct an orthogonal transformation matrix that performs a rotation of basis from the evaluation plane to the transversal plane. This matrix is shown in Equation \ref{eq:orthomatrix}

\begin{equation}
\centering
\mathbf{O} = 
    \begin{pmatrix}
     l_{\alpha} & l_{\beta} & l_{\gamma} \\
     m_{\alpha} & m_{\beta} & m_{\gamma} \\
     k_{\alpha} & k_{\beta} & k_{\gamma} \\
    \end{pmatrix},
    \label{eq:orthomatrix}
\end{equation}

Where $\mathbf{l}_{E}$,$\mathbf{m}_{E}$, and $\mathbf{k}_{cen,E}$, are written as $l,m,k$ respectively for brevity and are shown in Equation \ref{eq:orthomatrix} as their components projected onto the vectors $\boldsymbol{\alpha},\boldsymbol{\beta}$, and $\boldsymbol{\gamma}$. The matrix $\mathbf{O}$ is the tool we need to express our ray coordinates in terms of the transversal plane basis. This is done by a simple multiplication of the ray position and direction vectors ($\mathbf{r}'_{ray,E}$,$\mathbf{k}_{ray,E}$) with the matrix, shown in Equations \ref{eq:posbasis} and \ref{eq:dirbasis}

\begin{equation}
    \mathbf{r}_{ray,T} = \mathbf{O}\mathbf{r}'_{ray,E},
    \label{eq:posbasis}
\end{equation}

\begin{equation}
    \mathbf{k}_{ray,T} = \mathbf{O}\mathbf{k}_{ray,E}.
    \label{eq:dirbasis}
\end{equation}

The same must also be done for the field evaluation point of interest $\mathbf{r}_{point,E}$ to evaluate the Gaussian field later, shown in Equation \ref{eq:pixbasis},

\begin{equation}
    \mathbf{r}_{point,T} = \mathbf{O}\mathbf{r}_{point,E}.
    \label{eq:pixbasis}
\end{equation}.

Now that the ray data is expressed in the transversal plane coordinate system, we can use them to compute the differential ray transfer matrix used to propagate the Gaussian beamlet.

\subsection{Computing the differential ray transfer matrix}
To compute the differential ray transfer matrix for a Gaussian beam we require the five ray coordinates and directions derived in the previous subsection ($\mathbf{r}_{ray,T}$, and $\mathbf{k}_{ray,T}$), as well as the data for the same rays on the source plane ($\mathbf{r}_{ray,S}$, and $\mathbf{k}_{ray,S}$). The differential ray transfer matrix is calculated using the methods discussed in \hyperref[sec:methods]{Section \ref{sec:methods}}. Recall that an element of the ray transfer matrix is given by the difference of the central and differential ray coordinates at the source and evaluation plane (see Equation \ref{eq:Ayy}). The resultant matrix is given by Equation \ref{eq:diffmat}. The full matrix including the ray data used to compute each of the elements of Equation \ref{eq:diffmat}, can be found in \hyperref[sec:appendixA]{Appendix A}.

\begin{center}
\begin{equation}
    \begin{pmatrix}[c | c]
        \mathbf{A} & \mathbf{B} \\
        \hline
        \mathbf{C} & \mathbf{D} \\
    \end{pmatrix} 
    =
    \begin{pmatrix}[c c | c c ]
    \frac{\partial x_T}{\partial x_S} & \frac{\partial x_T}{\partial y_S} & \frac{\partial x_T}{\partial l_S} & \frac{\partial x_T}{\partial m_S}  \\
    \frac{\partial y_T}{\partial x_S} & \frac{\partial y_T}{\partial y_S} & \frac{\partial y_T}{\partial l_S} & \frac{\partial y_T}{\partial m_S}  \\
    \hline 
    \frac{\partial l_T}{\partial x_S} & \frac{\partial l_T}{\partial y_S} & \frac{\partial l_T}{\partial l_S} & \frac{\partial l_T}{\partial m_S} \\
    \frac{\partial m_T}{\partial x_S} & \frac{\partial m_T}{\partial y_S} & \frac{\partial m_T}{\partial l_S} & \frac{\partial m_T}{\partial m_S} \\
    \end{pmatrix}
    \label{eq:diffmat}
\end{equation}
\end{center}

The total ray transfer matrix can be organized into a tensor composed of 2x2 sub-matrices shown on the left hand side of Equation \ref{eq:diffmat}. These are exactly the non-orthogonal representations of the ray transfer matrix discussed in \hyperref[sec:methods]{Section \ref{sec:methods}} that we need to compute the propagated Gaussian field profile.

\subsection{Computing the Gaussian field}
The central ray serves as the coordinate origin for our Gaussian field evaluation on the transversal plane. With the position vectors effectively transformed in Equations \ref{eq:posbasis} and \ref{eq:pixbasis}, we can define our centered radial coordinate $\mathbf{r}_{o}$ as the simple difference of these two positions to center the coordinate system in Equation \ref{eq:radial_cord}
\begin{equation}
    \mathbf{r}_{o} = \mathbf{r}_{point,T} - \mathbf{r}_{cen,T}.
    \label{eq:radial_cord}
\end{equation}
Finally, we can call upon Equation \ref{eq:gaussian_prop} to perform the Gaussian field evaluation, with some added phase factors to account for the free-space propagation of the Gaussian beamlet. The result is shown in Equation \ref{eq:final_gauss}
\begin{equation}
    U(\mathbf{r}_{o}) = \frac{U_{o}}{\sqrt{det|\mathbf{A} + \mathbf{BQ_{1}^{-1}}|}} exp[\frac{-ik}{2} \mathbf{r}^{T}_{o} \mathbf{Q_{2}^{-1}} \mathbf{r}_{o} + ik\Delta_{cen} + ik\Phi_{opd}],
    \label{eq:final_gauss}
\end{equation}

where $\mathbf{Q_{2}^{-1}}$ is the propagated complex curvature matrix given by Equation \ref{eq:prop_complex}, $\mathbf{r}_{o}^{T}$ is the transpose of the coordinate from Equation \ref{eq:radial_cord}, $\Delta_{cen}$ is the propagated distance of the central ray to the transversal plane, and $\Phi_{opd}$ is the optical path experienced by the central ray through the optical system, which we get from the ray tracing engine used. Note that while the phase factors  $\Delta_{cen}$ and $\Phi_{opd}$ are not present in Equation \ref{eq:gaussian_prop}, they are necessary for GBD to correctly interfere all of the Gaussian beams that are propagated.

The procedure outlined above is repeated for each beamlet and location on the evaluation plane. From a computational perspective this procedure is somewhat daunting. Repeating this process means the computation complexity scales by the number of beamlets and number of points on the evaluation plane, which can quickly become prohibitive. Fortunately, the algorithm described is entirely vector and matrix operations so it is simple to take advantage of broadcasted array operations in Python to vectorize the computation, and parallel computing to accelerate it. In our implementation, the vector operations described above are broadcasted such that the field of a single beamlet at all points is computed simultaneously. This operation exists in a loop over the number of beamlets. For our preliminary efforts in computational acceleration, see \hyperref[sec:appendixB]{Appendix B}.

 Note the subtlety that the propagation procedure does not depend on the elementary field of choice until the electric field is evaluated. The assumption of this propagation method is that the elementary field is paraxial about the area encompassed by the differential rays. In principle, as long as a field's propagation formula is known analytically via the general Collins integral it can be propagated with this method. Thus, GBD is a \emph{special case} of the method described in this section. Because of the history of modeling resonators with ray transfer matrices, we know the ABCD propagation laws for Laguerre-Gaussian\cite{Mei_2005} and Hermite-Gaussian\cite{CAI2002139} beams, which form complete sets. This indicates that the beamlet decomposition method would be of higher accuracy when decomposing the field into modes of higher spatial order. The formula for flattened elliptical Gaussian beams is also known\cite{CAI2002139}, which are capable of mitigating the soft-edge effect imposed by the traditional beamlet decomposition\cite{greynolds_ten_2020}. This method also works with Worku and Gross's half- and quarter-truncated Gaussian beamlets\cite{Worku19}. The algorithm described in this Section is one of the key results of this study, because it can propagate the field of any known solution to the Collins integral to an arbitrary array of points. We build this algorithm into Poke to test its ability to mimic diffraction simulations with known results, and perform a comparison against commercial software. 

\section{Results}
\label{sec:results}
\begin{figure}[H]
    \centering
    \includegraphics[width=\textwidth]{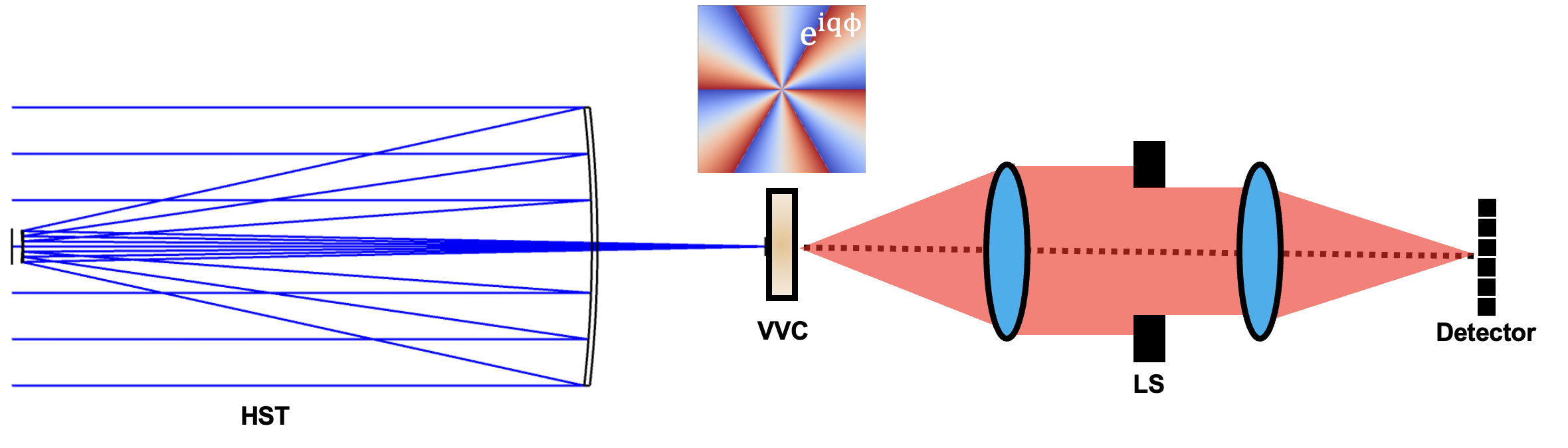}
    \caption{Illustration of the hybrid propagation physics model used to produce the results in this Section. The system prescription in Table \ref{tab:fiducial_observatory_specs} is loaded into Zemax OpticStudio and shown on the left (labeled HST). The phase of a vortex coronagraph (VVC) is shown in the middle. The GBD PSF is computed at this plane and propagated through the coronagraph using HCIPy to arrive at the final image at the detector plane.}
    \label{fig:telescope_model}
\end{figure}

To evaluate GBD as a viable physical optics propagation technique we benchmark its performance versus traditional diffraction simulations for a given observatory coupled to a vortex coronagraph. The fiducial observatory used in this study is a Ritchey-Chretien (RCs) objective based on the Hubble Space Telescope (HST) using an unobscured aperture. This model is constructed in Zemax OpticStudio, using the system prescription is given in Table \ref{tab:fiducial_observatory_specs} and is illustrated in Figure \ref{fig:telescope_model}.

\begin{table}[H]
    \centering
    \begin{tabular}{c c c c c}
        \hline
        Surface & RoC [m] & Conic Constant & Distance [m] & Diameter [m]  \\
        \hline
        M1 & -11.0400 & -1.00230 & -4.90607 & 2.40000 \\
        M2 & 1.35800 & -1.49686 &  6.40620 & 0.28112 \\
        \hline
        \\
    \end{tabular}
    \caption{Optical system prescription for the RC telescope based on the HST used in this investigation. All distances are given in meters. RoC stands for Radius of Curvature, and the sign convention is chosen such that negative values are concave, and positive values are convex.}
    \label{tab:fiducial_observatory_specs}
\end{table}

We first compare the PSFs generated by GBD to the analytical Airy function to assess the degree to which GBD can represent the focused field after a circular aperture. We then compare an aberrated GBD PSF to one produced by the Zemax Huygens PSF analysis tool to assess GBD's ability to reconstruct the field after a vignetted and aberrated wavefront. Finally, we propagate the focused field after a circular aperture produced by GBD through a Fraunhofer model of a vortex coronagraph and compare it to the results given by using solely Fraunhofer diffraction. The PSF simulations conducted in Section 4.2 are monochromatic simulations at 1.65 $\mu$m on a detector with 256 $\times$ 256 pixels over 1 $\times$ 1 mm (or 25 $\times$ 25 $\frac{\lambda}{D}$ ). The coronagraph simulations conducted in Section 4.3 are conducted at the same wavelength, but with 1600 $\times$ 1600 pixels across 8 $\times$ 8 mm (or 200 $\times$ 200 $\frac{\lambda}{D}$) to better sample the vortex mask and reach the desired contrast levels for both the Hybrid and Fraunhofer model.

\subsection{The Fiducial Coronagraph}
Our goal in this study is to assess the feasibility of GBD to integrate ray models of observatories into physical optics models of coronagraphs accurately. To quantify this, we propagate the images produced by GBD and the analytical Airy function through a charge-2 vortex coronagraph (VC) \cite{mawet_annular_2005,lee_experimental_2006}. The optical VC is the general case of the vector vortex coronagraph\cite{mawet_vector_2010} which has shown great promise for future missions to image Earthlike exoplanets \cite{serabyn_vector_2019-1}. These coronagraphs are excellent at rejecting low-order spatial modes, while transmitting the remainder of the light. We expect that traditional GBD will have some difficulty in accurately modeling high-spatial frequency content, and that it will manifest in the focal plane of this fiducial coronagraph if the error is limiting. If not, then we can conclude that GBD is suitable for high-contrast imaging simulation.

The complex amplitude of the charge-$q$ VC focal plane mask is given by
\begin{equation}
	U(x,y) = exp\Bigl[i q \arctan(\frac{y}{x})\Bigr],
	\label{eq:vortex}
\end{equation}
where $q$ is the topological charge and the transmission is unity everywhere except for the center pixel, where it is 0. This is because there is a singularity in the phase ramp at this location, so it must be masked out. We chose the VC because of its ability to effectively reject on-axis starlight at a given wavelength. Should GBD introduce undesirable artifacts into the PSF, it should be visible in the coronagraph focal plane. Modeling a VC accurately is challenging computationally, because the on-axis starlight is only completely rejected if the focal plane is infinitely sampled. The singularity at the center must also be sampled highly in order to accurately sample the rapid change in phase immediately around it without discretization errors. These require very large arrays ($\ge$ 16k $\times$ 16k arrays) for meaningful starlight rejection, which considerably slows the simulation. To overcome this computational burden, a multi-step propagation algorithm can be used to sample the central singularity higher than the rest of the field. HCIPy\cite{por2018hcipy} has this algorithm implemented in the \verb"VortexCoronagraph" class, which accepts a user-specified wavefront and then outputs the wavefront after the VC and before the Lyot stop. We can define our GBD PSF as an HCIPy wavefront and propagate it through the coronagraph to complete our hybrid propagation model and analyze the image plane residuals. We can then compare the residuals of our hybrid propagation model to one where the PSF of a circular aperture was computed with Fraunhofer diffraction. This permits us to compare the scale of the error introduced by the hybrid propagation model against the numerical simulation errors present in traditional diffraction simulation.

\label{sect:results}  

\subsection{The Observatory PSF}
\begin{figure}[H]
    \centering
    \includegraphics[width=\textwidth]{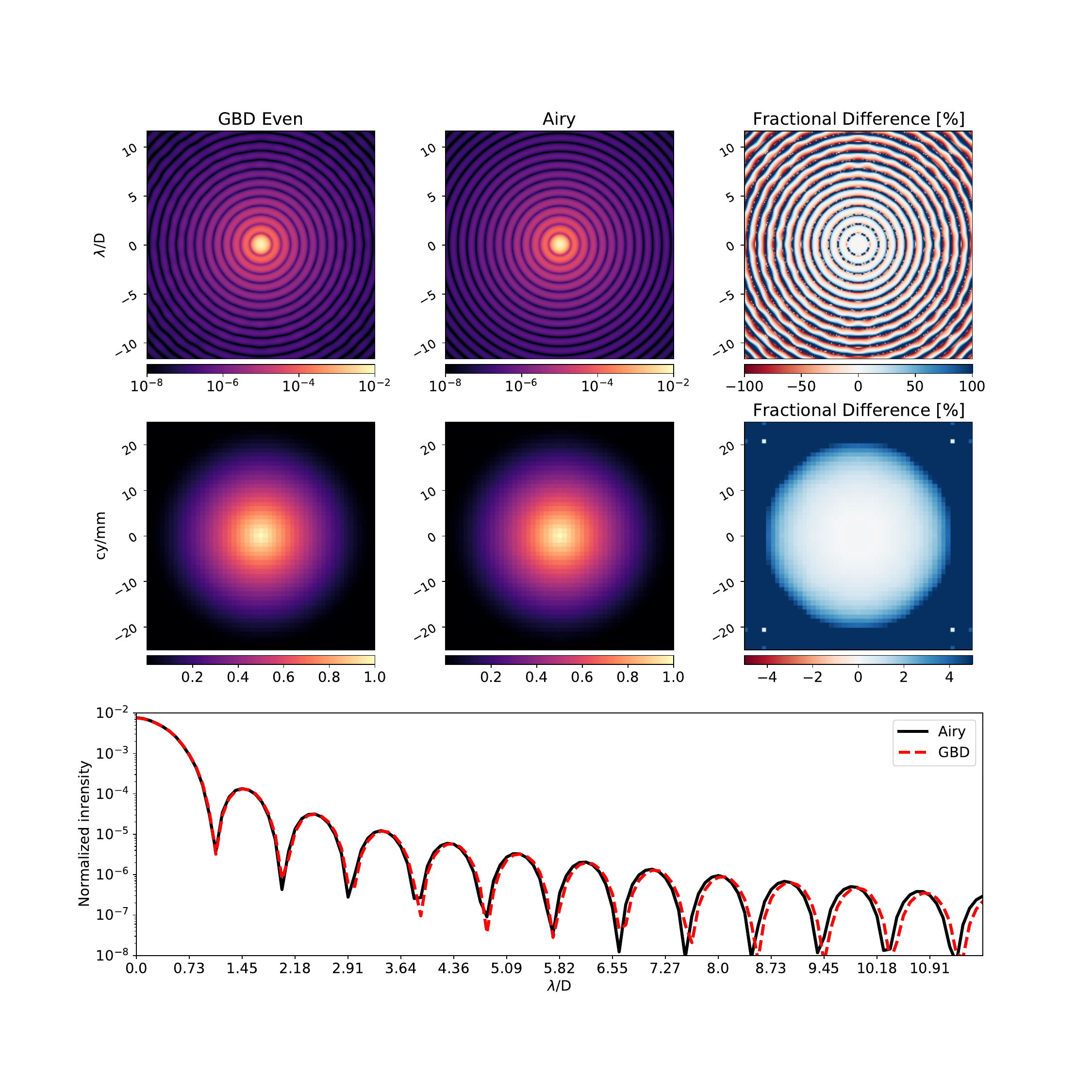}
    \caption{Comparisons of the PSF (top) and MTF (middle) for GBD with even sampling (left) and analytical Airy pattern (middle). The PSFs are given in units of normalized intensity, such that the sum of the energy in the PSF is unity. The fractional difference is plotted on the right, and the azimuthally averaged radial profile is plotted on the bottom. The MTF is plotted out to the cutoff frequency of the HST of around 25 cy/mm. The radial oscillation in the fractional difference PSF is indicative of the artifacts introduced by GBD. In the MTF it is apparent that frequencies below 20 cy/mm are well-maintained, but the higher spatial frequencies increase in fractional difference. The effect on the PSF is revealed in the radial profile, where the PSF appears to spread out at larger angular separations. The RMS difference of the PSF data is 2.3e-6.}
    \label{fig:airy_even}
\end{figure}

\begin{figure}[H]
    \centering
    \includegraphics[width=\textwidth]{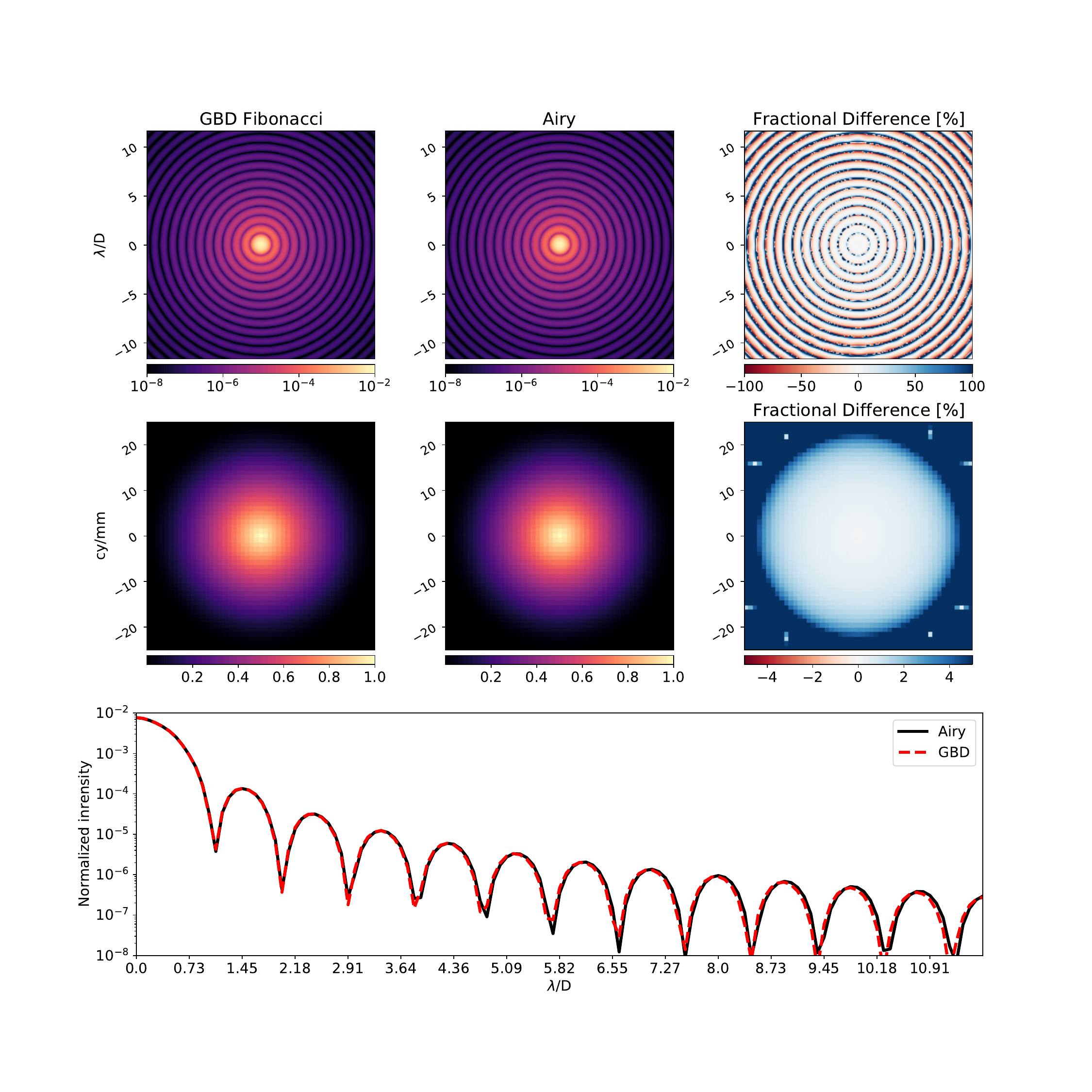}
    \caption{Comparisons of the PSF (top) and MTF (middle) for GBD with Fibonacci sampling for the same number of beamlets as Figure \ref{fig:airy_even} (left) and analytical airy pattern (middle). The PSFs are given in units of normalized intensity, such that the sum of the energy in the PSF is unity. The fractional difference is plotted on the right, with the MTF plotted out to the cutoff frequency of the HST of around 25 cy/mm. The azimuthally averaged radial profile is plotted on the bottom. The presence of ripples in the fractional difference PSF is noticeably lower than Figure \ref{fig:airy_even}, and the fractional difference MTF shows that frequencies out to 23 cy/mm are well-maintained. The spreading present in Figure \ref{fig:airy_even} is less prevalent in the radial profile. The RMS difference of the PSF data is 1.6e-6, indicating that this distribution results in a more accurate simulation.}
    \label{fig:airy_fib}
\end{figure}

First we examine GBD's ability to construct the Airy pattern. The Airy pattern represents the ``ideal" diffraction-limited observatory image for a circular aperture. The analytical solution for this image is known and available in POPPY, so we have a point of comparison that is not limited by numerical simulation errors. Shown in Figures \ref{fig:airy_even} and \ref{fig:airy_fib} are the PSF simulations generated by GBD using the even and Fibonacci sampling schemes. We also plot a comparison of the Modulation Transfer Function (MTF) to better illustrate how the transfer of individual spatial frequencies is affected by GBD. An important parameter in these simulations was the degree to which the effective diameter of the entrance pupil was appropriately captured. In Figures \ref{fig:gbd_essentials}c-d, we observe that GBD results in some energy spillover outside of the original aperture function, which would result in a PSF with an incorrect spatial extent. To mitigate this effect, we remove any beamlets within half a waist radius of the aperture boundary. This is a trade-off between too much energy outside of the aperture which results in a smaller PSF, and too little energy within the aperture which results in a larger PSF.

\begin{figure}[H]
    \centering
    \includegraphics[width=\textwidth]{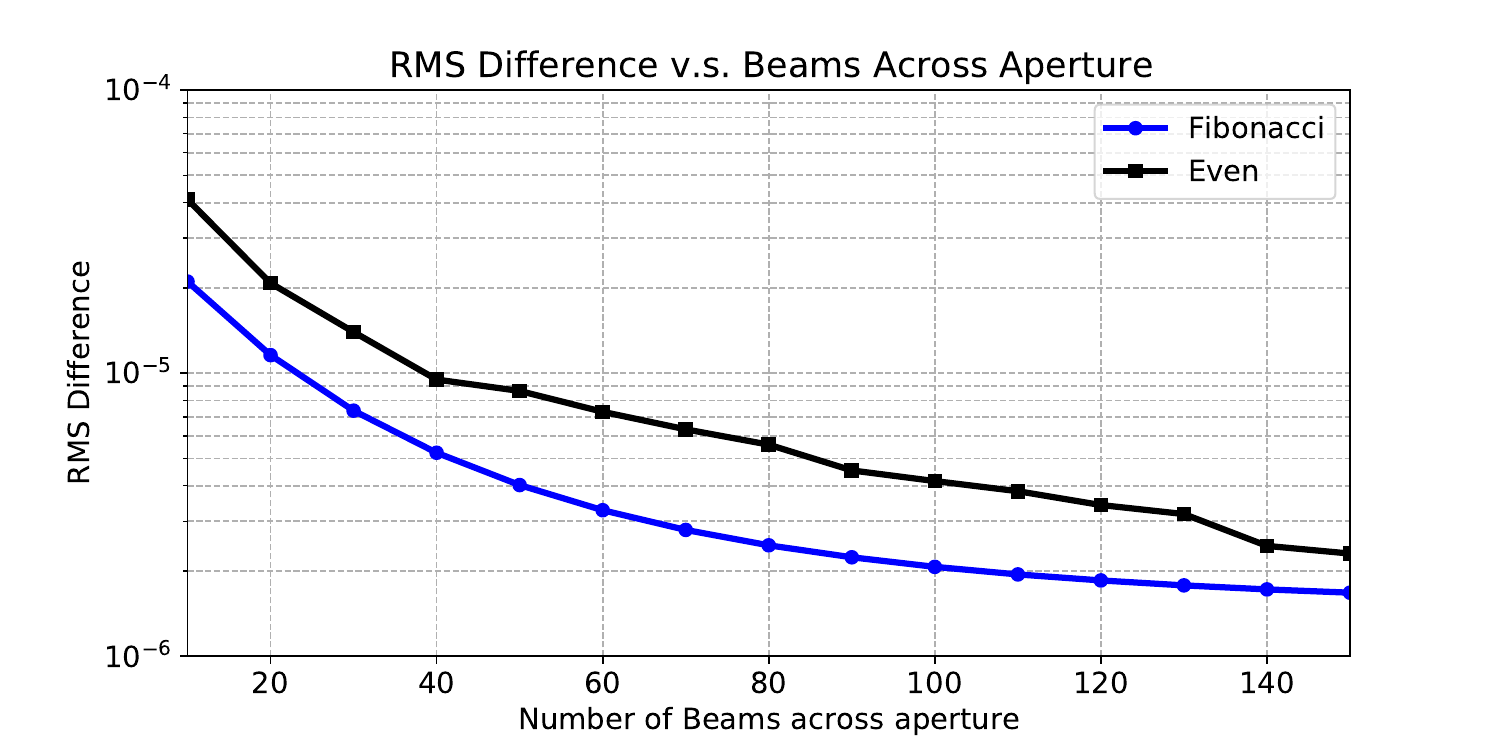}
    \caption{The RMS difference between the sum-normalized GBD PSF and analytical Airy function for the Fibonacci (blue circles) and Even (black squares) sample schemes. For the circular aperture it is clear that the Fibonacci sample scheme is more accurate for every case given a circular aperture. However, the returns are diminishing as the number of beamlets increases.} 
    \label{fig:rms_vs_sample_airy}
\end{figure}

Figures \ref{fig:airy_even} and \ref{fig:airy_fib} compare the same simulation of the Airy function where the only variable is how the entrance pupil was decomposed. In both the PSF and MTF dimensions it is clear that the Fibonacci sampling is the superior decomposition method for this aperture. For more under-sampled cases there may be a tradeoff in accuracy. To understand this, in Figure \ref{fig:rms_vs_sample_airy} we examine how well the analytical Airy function is reconstructed for the two sample schemes as a function of the number of beamlets.

The Fibonacci sampling clearly wins out in terms of performance given a fixed number of beamlets, which translates directly to computation time. This also means that using the Fibonacci sample scheme can yield a simulation of the same accuracy as the even sample scheme with fewer beamlets. By judicious choice in sample scheme, the computational complexity of GBD can be lessened or the simulation accuracy can be increased.

\begin{figure}[H]
    \centering
    \includegraphics[width=0.9\textwidth]{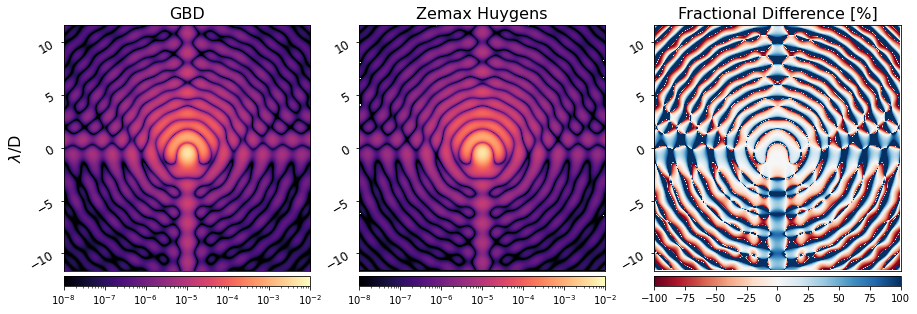}
    \caption{Simulations of an aberrated PSF with secondary support structure in the entrance pupil of our HST model given in Table \ref{tab:fiducial_observatory_specs}. The GBD PSF (left) and Zemax Huygens PSF (middle) have very similar structure, with the dominant difference in structure being the features from the sharp-edge diffraction from the spiders (shown in fractional difference on the right). This simulation helps quantify GBD's difficulty in modeling sharp-edge diffraction effects. The RMS difference of the PSF data is 7.5e-6.}
    \label{fig:aberrated}
\end{figure}

To further test GBD's capability to model a more realistic observatory PSF, we add obscurations in the entrance pupil that correspond to the secondary mirror and supporting spiders. We also tilt the secondary mirror by 0.05 degrees to aberrate the beam. By doing so, we simultaneously test our algorithm's ability to capture aberrations from optical misalignment and diffraction from structure in the aperture. The results of this simulation are shown in Figure \ref{fig:aberrated}.

The data in Figure \ref{fig:aberrated} was simulated in a comparison between our proposed algorithm and Zemax's Huygens PSF analysis tool. The Huygens PSF is computed by propagating spherical waves along ray paths in Zemax to the image plane and coherently summing them. The Zemax documentation asserts that this method makes fewer assumptions than the FFT PSF, so it was chosen as the point of comparison for the aberrated PSF. In our GBD simulation we vignette any Gaussian Beam that has any of its differential rays vignetted. Our results show extremely similar structure in the PSF from both the aberration induced by the secondary mirror misalignment and the structure from the HST spiders. The fractional difference reveals that the first couple ``rings" of the PSFs are almost identical, indicating that the low-order aberrations were sufficiently simulated by GBD. There is a larger fractional difference in the dimmer structures that result from the sharp edges of the secondary obscuration and spiders, which is a known challenge for GBD simulations. This result can be further improved by careful implementation of alternative beamlet profiles, such as Worku and Gross's truncated beamlets\cite{Worku19}, or higher-order transverse electric field modes (e.g. Hermite, Laguerre-Gaussian). However, the result using fundamental Gaussian modes is encouraging, since the RMS difference of the PSFs is within the same order of magnitude as the results shown in Figures \ref{fig:airy_even} and \ref{fig:airy_fib}. Zemax is an industry standard optical modeling tool, but its propagation algorithms are closed-source. It is beyond the scope of this paper to validate the accuracy of Zemax's PSF simulation tools, but it is encouraging that our results in Figure \ref{fig:aberrated} agree. Now that our algorithm's ability to perform PSF simulations has been verified, we can analyze the degree to which it introduces artifacts in high-contrast imaging simulations.

\subsection{Coronagraph Response}
\begin{figure}[H]
    \centering
    \includegraphics[width=0.9\textwidth, trim={3cm 0cm 2cm 0cm}]{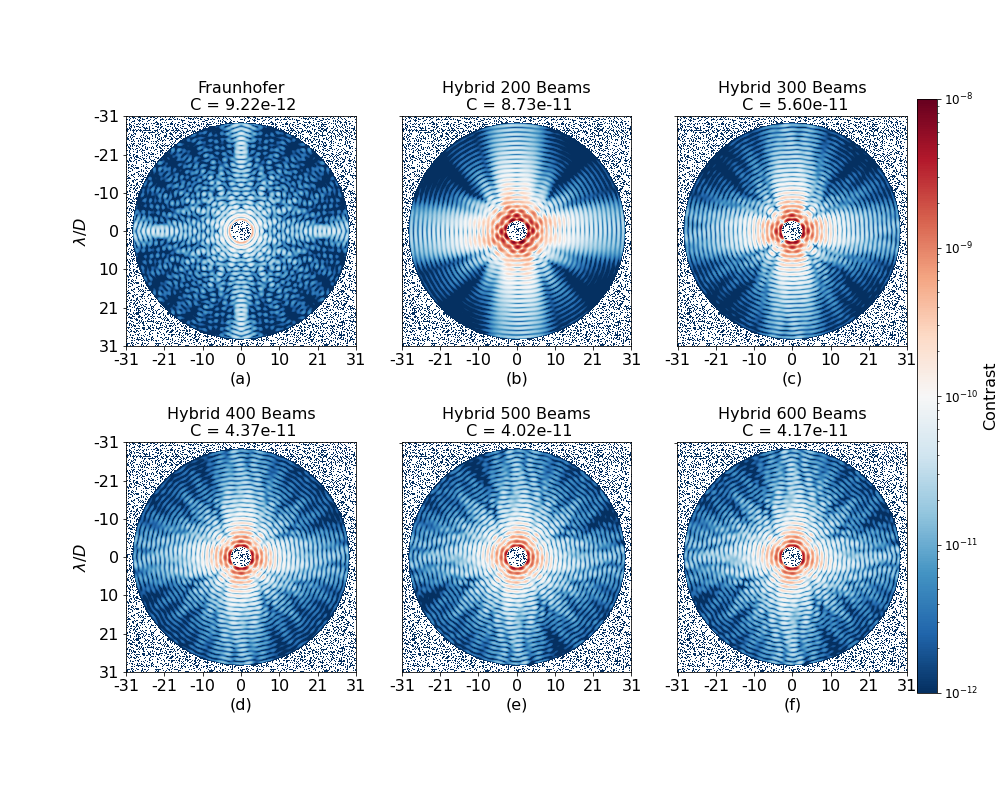}
    \caption{Comparison of the coronagraph PSF from 3-30 $\lambda / D$ generated by a solely Fraunhofer diffraction model (a) and the proposed Hybrid propagation scheme for a varying number of beamlets across the 2.4m unobscured HST aperture with an overlap factor of 2. (b) was generated with 200 beamlets, (c) with 300 beamlets, (d) with 400 beamlets, (e) with 500 beamlets, and (f) with 600 beamlets across the aperture. These data were generated by producing a PSF with the given propagation scheme and then propagating it through HCIPy's VortexCoronagraph class with a topological charge of 2. To reflect the residual starlight's astrophysical flux ratio ``contrast", the PSF data are normalized to the maximum of an off-axis PSF at $\approx 16 \lambda/D$ that propagates through the VortexCoronagraph. The mean contrast of the masked region (C) is shown on each image. The residuals from the Hybrid propagation (b-f) are brighter than the equivalent Fraunhofer simulation, and minimize at the 500 beamlet case. This is particularly apparent near the inner working angle, where we see a maximum signal of $\approx$ $1 \times 10^{-8}$ for the 200 beamlet case decrease to $\approx 5 \times 10^{-9}$ in the 500 beamlet case.}
    \label{fig:nonparaxial_coronagraph}
\end{figure}

In traditional VCs, the on-axis field from an unvignetted circular aperture should be entirely rejected. We expect from the observatory PSF simulations that GBD does not trace high-spatial frequency information well due to the soft edges and amplitude ripples introduced by the beamlet decomposition. Any meaningful errors from this step should pass through the coronagraph unperturbed. To formally assess GBD's suitability for high-contrast imaging, we construct the PSF of the fiducial observatory with a circular aperture using GBD and compare it to a Fraunhofer model of the same system. Both PSFs are propagated with Fraunhofer diffraction through the vortex coronagraph and the coronagraph focal plane is compared to assess the presence, if any, of residual signal introduced by GBD. The parameters used in this simulation are given in Table \ref{tab:coro_params}, and the results shown in Figure \ref{fig:nonparaxial_coronagraph} are cropped to show the innermost 30 $\lambda / D$ to better display the structure near the inner working angle.

\begin{table}[H]
    \centering
    \begin{tabular}{c|c}
       \hline
       Parameter  & Value  \\
       \hline
        Wavelength ($\lambda$) & 1.65 $\mu$m \\
        Number of Pixels ($N_{pixels}$) & 1600 \\
        Instantaneous FoV ($\Delta x$) & 4.95 $\mu$m or $\frac{1}{8}\frac{\lambda}{D}$\\
       \hline
       \hline
    \end{tabular}
    \\
    \caption{Simulation parameters for the result shown in Figure \ref{fig:nonparaxial_coronagraph}. 
    }
    \label{tab:coro_params}
\end{table}

Figure \ref{fig:nonparaxial_coronagraph} shows the residuals that propagate through to the final image plane which arise from inaccuracies in the propagation model. The Hybrid propagation model generally minimizes in average contrast until the 600 beamlet case (Figure \ref{fig:nonparaxial_coronagraph}f) where there is a slight increase. The residuals from traditional Fraunhofer diffraction leave some low-energy features near the core of the PSF, which can be improved with greater sampling but the rest of the field is largely below $10^{-11}$ contrast. The decreasing residual energy with the number of beamlets used in the hybrid propagation scheme is encouraging, but we appear to have found the point at which increasing the number of Gaussian beams no longer increases simulation accuracy. 

The algorithm outlined in \hyperref[sec:algorithm]{Section \ref{sec:algorithm}} is carried out as-written with some computational acceleration done by taking advantage of Python's ability to vectorize matrix operations and other Python packages to accelerate the exponential calculation, discussed in detail in \hyperref[sec:appendixB]{Appendix B}. We have not formally explored parallel processing packages on Central Processing Units (CPUs) or Graphical Processing Units (GPUs) to accelerate this computation, but expect that these could make higher-sampled simulations more feasible to minimize the artifacts that remain in GBD PSFs for high-contrast imaging simulations.
Now that we have established an open-source platform for GBD, Worku's Modified GBD\cite{Worku19} could be developed to minimize the number of beamlets required to minimize the artifacts in the coronagraphic focal plane. Exploring higher order spatial modes and the astigmatic fundamental mode of Gaussian Beams could also yield greater accuracy for the same or less computational complexity. 
The result in Figure \ref{fig:nonparaxial_coronagraph} using traditional GBD shows that we can reduce the residuals to below $10^{-9}$ Contrast near the inner working angle. It is also worth noting that this result suggests that with sufficient sampling, GBD can presently be used to simulate the PSFs for systems with less stringent contrast floors. To minimize the residuals from the propagation technique, other decomposition methods must be explored. We have created a platform for the algorithm's development in an open-source environment and will outline a road map for the most pressing optimizations that can be conducted by future investigations to improve GBD's accuracy.


\section{Summary and Conclusions}
\label{sec:conclusion}  

Diffraction-limited optical systems require an accurate physical optics model to simulate the performance of the instrument. We formally describe an alternative implementation of the Gaussian Beamlet Decomposition physical optics propagation technique that is well-suited to PSF simulation. We illustrate the degree to which GBD is capable of accurately modeling the PSF of a Hubble-like astronomical observatory, and quantify the artifacts that remain in the ``hybrid" simulation of a vortex coronagraph. We have demonstrated a new means of integrated observatory modeling which reaches near state-of-the-art contrast levels at $< 10^{-9}$ with an numerical average contrast of $4\times10^{-11}$.
To our knowledge this manuscript is the first of its kind to explicitly publish the full transfer matrix GBD method and evaluate its ability in the context of astronomical telescopes. In doing so we have made public a method of ray-based physical optics that has the potential to further integrate the optical modeling pipeline.

\subsection{Future Work}
The simulations presented in this manuscript, while accurate, necessitated highly-sampled simulations that were computationally intensive. The longest simulation used as many as 360,000 Gaussian beamlets across 2,560,000 pixels, which creates a phase array that is approximately 100TB in size. This simulation was done on an AMD Ryzen9 3950X processor and took $\approx$ 24 hours to complete. 
For GBD to be a practical diffraction technique for high-contrast instrument sensitivity analysis it would be ideal to generate a field of the same accuracy in a shorter amount of time. Preliminary efforts in accelerated computing are discussed in \hyperref[sec:appendixB]{Appendix B}, where we were able to reduce simulation runtime from taking $\approx$5 hours to complete, to $\approx$10 minutes. We intend to submit a follow-up study that outlines a more memory-efficient algorithm that we can use to more rapidly explore the degrees of freedom available in GBD (e.g. number of beamlets, overlap factor, etc.) for optimal diffraction simulation.

Modified GBD is already known to increase simulation accuracy for fewer beamlets\cite{Worku19}, and as such is the next natural step in the development of our GBD module. Worku and Gross's work on truncated Gaussian beams showed high accuracy in reconstructing the field after 2D polygonal apertures, and ``squeezed" the half-truncated beamlets in the azimuthal direction to increase the accuracy of a field after a circular aperture. Hexagonal apertures are of particular interest to astronomy because of their use in telescopes such as the W.M. Keck Observatory and James Webb Space Telescope. Understanding how the truncated beamlets are able to reconstruct the field after segmented apertures is another important step in developing GBD for high-contrast imaging simulations. It is also worth repeating the notion in \hyperref[sec:algorithm]{Section \ref{sec:algorithm}} that nothing about the proposed algorithm requires the beamlets to be strictly Gaussian. Therefore, considering alternative beamlets to use in the field decomposition may increase the accuracy beyond what traditional GBD and modified GBD are capable of. GBD has previously demonstrated the ability to model plane-to-plane diffraction effects (e.g. the spot of Arago)\cite{Harvey15}, and its ability to model diffraction from surface polishing errors should be investigated for a more comprehensive modeling pipeline.

The beamlet decomposition algorithm would also benefit from more consideration into leveraging how parallelizeable it is. The contribution from each beamlet at each pixel is computed independently. With thoughtful consideration to the structure of our code and wide library of parallel processing packages available to us in Python on CPU's and GPUs\cite{lam_numba_2015,robert_mcleod_2018_2483274}, the beamlet computation could be rapidly accelerated. Preliminary experiments on GPUs have shown a decrease in runtime by a factor of $\approx$16x. Several investigators have been exploring the Jax Python package for its ability to perform automatic differentiation and parallel computing in support of physical optics simulations\cite{Desdoigts2022,Wong21,Pope21}. Automatic differentiation could be extremely useful for future beamlet decomposition algorithms to improve the accuracy of the ABCD matrix computation. Parallelization and vectorization is the most natural path forward for our beamlet decomposition algorithm because of the independence of the beamlet operations. Jax makes this process simple with the \emph{pmap} and \emph{vmap} functions, while also allowing for just-in-time compilation for accelerated computing. 

The use of a beamlet decomposition algorithm completely integrates a diffraction model with a ray model of an optical system, resulting in a more physically complete  modeling pipeline. Consequently, other ray-based analyses can be integrated directly into the diffraction model. Polarization ray tracing\cite{Chippman15} is a natural extension to beamlet decomposition simulations because it can trace the complex amplitude of individual beamlets for generally vectorial field propagation through optical systems. This capability was demonstrated by Worku and Gross\cite{Worku17} in the context of high-numerical aperture microscope objectives, but has the potential to be a powerful simulation tool for astronomical telescopes, including the next generation giant segmented mirror telescopes (Thirty Meter Telescope, Extremely Large Telescope, Giant Magellan Telescope)\cite{anche_inprep} and the Astro2020-recommended IROUV space observatory, which may be sensitive to polarization aberrations.

\subsection{Open-Source Science and Engineering}
This research was inspired by POPPY, a Python physical optics module originally developed to simulate the James Webb Space Telescope. Our goal is to expand the capabilities of POPPY by investigating new propagation physics. We are developing the GBD module in a self-contained package with interfaces to other popular diffraction codes (POPPY, HCIPy) to provide ray information to diffraction models. The code used to prototype the GBD method can be found in the Poke repository on GitHub\cite{Ashcraft_poke_2022}. This repository is intended to be purely experimental as we develop the propagation physics for other high-contrast imaging packages that have substantive support. 

\section{Appendix A: Full Differential Ray Transfer Matrix Calculation}
\label{sec:appendixA}

Equation \ref{eq:total_abcd_matrix} contains the explicit ray data used in the computation of each element of the differential ray transfer matrix. Each element of the matrix is given by the form $a_{ray,B}$ where $a$ denotes the data (position $x,y$ or direction cosine $l,m$), $ray$ denotes which of the 5 rays the data is from ($+x$,$+y$,$+l$,$+m$,$cen$), and $B$ which denotes the plane the data was taken from ($S$ for source plane, $T$ for transversal plane).

\label{sec:appc}
\begin{equation}
\centering
    \begin{bmatrix}[c | c]
        \mathbf{A} & \mathbf{B} \\
        \hline
        \mathbf{C} & \mathbf{D} \\
    \end{bmatrix} =
    \renewcommand\arraystretch{1}
    \begin{pmatrix}[c c | c c]
        \frac{x_{+x,T} - x_{cen,T}}{x_{+x,S} - x_{cen,S}} & \frac{x_{+y,T} - x_{cen,T}}{y_{+y,S} - y_{cen,S}} & \frac{x_{+l,T} - x_{cen,T}}{l_{+l,S} - l_{cen,S}} & 
        \frac{x_{+m,T} - x_{cen,T}}{m_{+m,S} - m_{cen,S}} \\
        \frac{y_{+x,T} - y_{cen,T}}{x_{+x,S} - x_{cen,S}} & \frac{y_{+y,T} - y_{cen,T}}{y_{+y,S} - y_{cen,S}} & \frac{y_{+l,T} - y_{cen,T}}{l_{+l,S} - l_{cen,S}} & 
        \frac{y_{+m,T} - y_{cen,T}}{m_{+m,S} - m_{cen,S}} \\
        \hline 
        \frac{l_{+x,T} - l_{cen,T}}{x_{+x,S} - x_{cen,S}} & \frac{l_{+y,T} - l_{cen,T}}{y_{+y,S} - y_{cen,S}} & \frac{l_{+l,T} - l_{cen,T}}{l_{+l,S} - l_{cen,S}} & 
        \frac{l_{+m,T} - l_{cen,T}}{m_{+m,S} - m_{cen,S}}\\
        \frac{m_{+x,T} - m_{cen,T}}{x_{+x,S} - x_{cen,S}} & \frac{m_{+y,T} - m_{cen,T}}{y_{+y,S} - y_{cen,S}} & \frac{m_{+l,T} - m_{cen,T}}{l_{+l,S} - l_{cen,S}} & 
        \frac{m_{+m,T} - m_{cen,T}}{m_{+m,S} - m_{cen,S}} \\
    \end{pmatrix}
\label{eq:total_abcd_matrix}
\end{equation}

\section{Appendix B: Accelerated Computing}
\label{sec:appendixB}
The independence of the Gaussian beamlet operations are uniquely suited to the exploration of multi-threaded computation to accelerate diffraction simulations. Accelerated computing is integral to diffraction modeling to enable rapid and precision simulation of small signals. 
The time to conduct traditional Fourier-based diffraction modeling is set by the complexity of the system. The sampling of each optical element and the number of total optical elements increase the complexity and number of Fast Fourier Transforms (FFTs) used, resulting in more computation time. GBD circumvents the FFT entirely by tracing rays to propagate through the optical system in a fraction of the time of the FFT. GBD’s diffraction calculation at the plane of interest computes an exponential of a complex-valued array that scales with the number of beamlets and sampling of the image plane, resulting in longer computation times.  Preliminary explorations into accelerated computing were conducted using the numexpr\cite{robert_mcleod_2018_2483274} and numba\cite{lam_numba_2015} Python packages that showed favorable computation time decreases by multi-threading the operation on a CPU. Both packages operate by pre-compiling a given function into machine code that the program calls and breaking up the operation into chunks of arrays that a CPU core can handle efficiently. The key computational advantage of GBD is the ability to do diffraction calculations in parallel. Numba was the first package explored due to its ease of implementation. The package works by applying a decorator to a Python function that processes the large array of interest, and then specifying the number of central processing unit (CPU) cores for the process to use. The distribution of the information stored within the array is handled automatically by numba, and results in considerable runtime decreases for GBD. On a 16 core 2.4GHz CPU, runtime for a simulation of 1876 Gaussian beamlets through a coronagraph to simulate a 256x256 pixel focal plane was sped up by a factor of 5, which approached POPPY's Fresnel diffraction runtime. This experiment was repeated using the numexpr package which showed an even greater decrease in computation time, consistent with results for accelerating Fresnel diffraction in POPPY\cite{Doug18}. The comparison in runtime vs. number of CPU cores is shown in Figure 6.

\begin{figure}[H]
    \centering
    \includegraphics[width=0.7\textwidth]{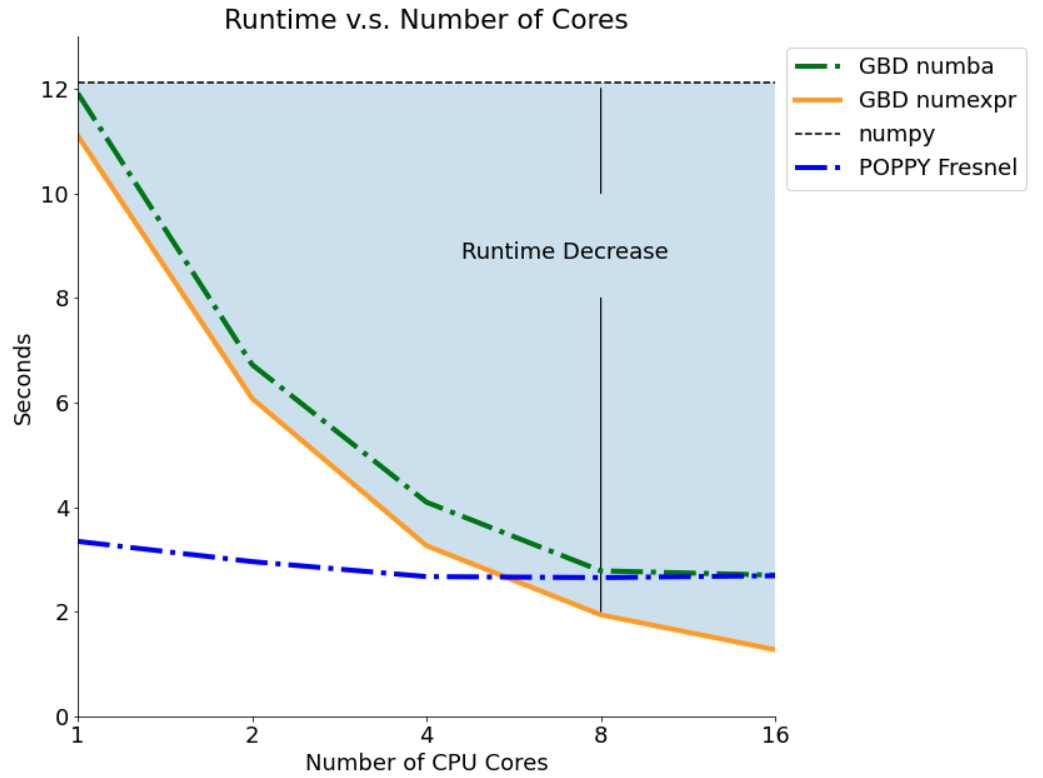}
    \caption{Run time comparison for a 50x50 grid of Gaussian beamlets on a 256 x 256 detector grid v.s. number of CPU cores used. Numexpr considerably accelerates the exponential calculation given multiple processing cores, and is an encouraging path forward for further parallelization.}
    \label{fig:runtime_sim}
\end{figure}

\begin{figure}[H]
    \centering
    \includegraphics[width=0.7\textwidth]{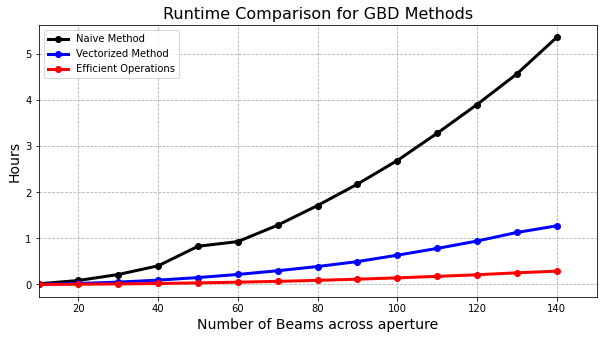}
    \caption{Runtime comparison of the iterations of the GBD algorithm applied in Python 3.8. The version using nested for loops is plotted in black, and high-beamlet simulations were computationally prohibitive. The runtime for the vectorized code is shown in blue, and the vectorized code with more efficient linear algebra operations (determinant, inverse) is shown in red. Overall, the decrease in runtime from the black curve to the red curve is $\approx 40\times$, which enabled the highly-sampled simulations in Figure \ref{fig:nonparaxial_coronagraph}.}
    \label{fig:runtime_compare}
\end{figure}

The vector operations used in the algorithm described in \hyperref[sec:algorithm]{Section \ref{sec:algorithm}} can be broadcasted across the entire array, enabling more efficient computation of the ABCD matrix. Broadcasting refers to the practice of writing code such that a given operation is applied to each element of the array simultaneously. This is more commonly known as "vectorized" computing. Our initial demonstration of the GBD algorithm was written using nested for loops (black, Figure \ref{fig:runtime_compare}), which was inefficient for highly-sampled simulations. Vectorizing our code resulted in a $\approx10\times$ decrease in runtime (blue, Figure \ref{fig:runtime_compare}). We then determined that the elementary matrix operations done using Numpy's linear algebra library were considerably dominating the runtime, so we wrote our own versions of the determinant, and inverse functions. This change resulted in a net $\approx40\times$ decrease from the original code written using nested loops (red, Figure \ref{fig:runtime_compare}). The runtimes for a simulation of various numbers of beams across the aperture to a 256$\times$256 pixel detector are shown in Figure \ref{fig:runtime_compare}.

We anticipate even greater speedups on GPUs. In particular, the creation of the phase array from the ray data is the slowest operation and needs to be parallelized for accelerated diffraction modeling. Using the Cupy python package, adding support for GPUs was made quite simple given its compatibility with the numpy API. Preliminary experiments using our beamlet decomposition algorithm show a $\approx$16$\times$ decrease in computation time versus the algorithm shown in red on Figure \ref{fig:runtime_compare}. We have added GPU support though Cupy in our Poke package and plan to publish a formal study quantifying the degree to which GBD can benefit from accelerated computing in a future report.

\section{Code, Data, and Materials Availability Statement}

The code used to conduct these simulations is located on the Poke repository on GitHub in the \verb"GBD_Paper" directory. The \verb"test_worku_transversal.py" script was used to conduct the simulations in this manuscript. These simulations were done during a very early release (v0.1.0) of Poke without much optimization for runtime or a user-friendly design. This version has been moved to the legacy branch of the repository as Poke and the GBD module is actively developed. Presently, the updated version of this algorithm exists in Poke v1.1.0 as the poke.gbd module.



\bibliography{report,ewan_bib}   
\bibliographystyle{spiejour}   

\section{Acknowledgments}
This work benefited greatly from the insight of Dr. Norman Girma Worku, who pioneered the development of the transfer-matrix method of Gaussian Beamlet Decomposition. We thank Trenton Brendel, Weslin Pulin, and Brandon Dube for helpful discussion on the fundamental physics and design of the Poke API. Thanks to Kian Milani for the guidance with matplotlib. Thanks to Marcos Esparza, Kevin Derby, Hyukmo Kang, and Ramya Anche for help proof-reading this manuscript. Thanks to Sebastiaan Haffert for guidance on the use of HCIPy's VortexCoronagraph model. This research made use of several open-source Python packages, including POPPY\cite{Perrin12}, HCIPy\cite{por2018hcipy}, prsym\cite{Dube2019}, numpy\cite{harris2020array}, matplotlib\cite{Hunter:2007}, ipython\cite{PER-GRA:2007}, astropy\footnote{http://www.astropy.org} \cite{astropy:2013, astropy:2018, astropy:2022},  scipy\cite{2020SciPy-NMeth}, numba\cite{lam_numba_2015}, and numexpr\cite{robert_mcleod_2018_2483274}. This research made use of High Performance Computing (HPC) resources supported by the University of Arizona TRIF (Technology and Research Initiative Fund), UITS, and Research, Innovation, and Impact (RII) and maintained by the UArizona Research Technologies department. This work was supported by a NASA Space Technology Graduate Research Opportunity.


\vspace{2ex}\noindent\textbf{Jaren N. Ashcraft} (He/They) is a Ph.D. Candidate at the University of Arizona's Wyant College of Optical Sciences working with the UA Space Astrophysics Laboratory and Large Optics Fabrication and Testing Group. He received his B.S. degree in Optical Engineering from the University of Rochester in 2019, and M.S. degree in Optical Sciences from the University of Arizona in 2022. He is a recipient of a NASA Space Technology Graduate Research Opportunities award.

\vspace{2ex}\noindent\textbf{Ewan S. Douglas (He/they)}  is an Assistant Professor of Astronomy at the University of Arizona, and an Assistant Astronomer, Steward Observatory. Ewan completed a postdoctoral appointment at the Massachusetts Institute of Technology Department of Aeronautics and Astronautics, received master's and doctoral degrees in Astronomy from Boston University, and a bachelor's degree in physics from Tufts University.  Dr. Douglas' specializes in space telescopes and instrumentation for high-contrast imaging of debris disks and extrasolar planets.

\vspace{2ex}\noindent\textbf{Daewook Kim} is a faculty of optical sciences and astronomy at the University of Arizona. His research area covers precision optical engineering, optics fabrication, and freeform metrology including interferometry and deflectometry. He is the chair of the Optical Manufacturing and Testing (SPIE) and Optical Fabrication and Testing (OPTICA) conferences. He has served as an associate editor for the Optics Express journal. He is a senior member of OPTICA and SPIE Fellow.

\vspace{2ex}\noindent\textbf{Dr. A J Eldorado Riggs} is an optical engineer at the Jet Propulsion Laboratory. His research focuses on the high-contrast imaging of exoplanets, in particular mask optimization and wavefront sensing and control for the Coronagraph Instrument on the Nancy Grace Roman Space Telescope. He received his B.S. in physics and mechanical engineering from Yale University in 2011 and his Ph.D. in mechanical and aerospace engineering from Princeton University in 2016.

\vspace{1ex}


\end{spacing}
\end{document}